\documentclass[a4paper]{article}
\usepackage{amsmath,amssymb}

\DeclareMathOperator{\diag}{diag}
\newcommand{\nad}[1]{\mbox{\smash{\oalign{$#1$ \crcr \hidewidth 
$\mathchar"017E$ \hidewidth}}}}

\begin{document}
\bibliographystyle{unsrt}

\title{Tachyons and the preferred frames\thanks{\emph{Int.\ J. Mod.\ Phys.\ 
A}, to appear.}}
\author{Jakub Rembieli\'nski\thanks{\textit{E-mail address}: 
jaremb@mvii.uni.lodz.pl}\\ Katedra Fizyki Teoretycznej, 
Uniwersytet {\L}\'odzki\\ ul.~Pomorska 149/153, 90--236 {\L}\'od\'z, Poland}
\date{}
\maketitle

\begin{abstract}
Quantum field theory of space-like particles is investigated in the 
framework of absolute causality scheme preserving Lorentz symmetry. 
It is related to an appropriate choice of the synchronization procedure
(definition of time). In this formulation existence of field excitations 
(tachyons) distinguishes an inertial frame (privileged frame of reference) 
\emph{via} spontaneous breaking of the so called synchronization group. In 
this scheme relativity principle is broken but Lorentz symmetry is exactly 
preserved in agreement with local properties of the observed world.
It is shown that tachyons are associated with unitary orbits of Poincar\'e 
mappings induced from $SO(2)$ little group instead of $SO(2,1)$ one. 
Therefore the corresponding elementary states are labelled by helicity. The 
cases of the helicity $\lambda = 0$ and $\lambda = \pm\frac{1}{2}$ are 
investigated in detail and a corresponding consistent field theory is 
proposed. In particular, it is shown that the Dirac-like equation proposed 
by Chodos \textit{et al}. \cite{CHK}, inconsistent in the standard 
formulation of QFT, can be consistently quantized in the presented 
framework. This allows us to treat more seriously possibility that neutrinos 
might be fermionic tachyons as it is suggested by experimental data about 
neutrino masses \cite{PDG,Ass,GR}.
\end{abstract}

\section{Introduction}
Almost all recent experiments, measuring directly or indirectly the electron 
and muon neutrino masses, have yielded negative values for the mass 
square\footnote{A direct measurement of the electron neutrino mass is made 
in several tritium beta decay experiments. They are sensitive to a small 
neutrino mass because the energy release of the decay is small. A 
systematic theoretical analysis of this process \cite{Bro} shows that 
possible corrections to experimental result does not suffice to explain the 
mysterious negative mass square.} \cite{PDG,Ass,GR}. It suggests that these 
particles might be fermionic tachyons. This intriguing possibility was 
written down some years ago by Chodos \emph{et al}.\ \cite{CHK} and Recami 
\emph{et al}.\ \cite{GMMR}.

On the other hand, in the current opinion, there is no satisfactory theory 
of superluminal particles; especially it is commonly believed that there is 
no respectable tachyonic quantum field theory at present \cite{KK2}. This 
persuasion creates a psychological barrier to take such possibility 
seriously. Even if we consider eventuality that neutrinos are tachyons, the 
next problem arises; namely a modification of the theory of electro-weak 
interaction will be necessary in such a case. But, as we known, in the 
standard formulation of special relativity, the unitary representations of 
the Poincar\'e group, describing fermionic tachyons, are induced from 
infinite dimensional unitary representations of the non-compact $SO(2,1)$ 
little group.  Consequently, in the conventional approach, the neutrino 
field should be infinite-component one so a construction of an acceptable 
local interaction is extremally difficult.

In this paper we suggest a solution to the above dilemma. To do this we 
use the ideas developed in the papers \cite{Rem:tac,Rem:neu} based on the 
earlier works \cite{Rem1,Rem2}, where it was proposed a consistent 
description of tachyons on both classical and quantum level. The basic idea 
is to extend the notion of causality without a serious change of special 
relativity. This can be done by means of a freedom in the determination of 
the notion of the one-way light velocity, known as the ``conventionality 
thesis'' \cite{Rei,Jam}.

The main results of this paper can be summarized as follows:
\begin{itemize}
\item The relativity principle and the Lorentz covariance are formulated in 
the framework of a nonstandard synchronization scheme (the 
Chang--Thangherlini (CT) scheme). The absolute causality holds for all kinds 
of events (time-like, light-like, space-like).
\item For bradyons and luxons our scheme is fully equivalent to the standard 
formulation of special relativity. 
\item For tachyons it is possible to formulate covariantly canonical 
formalism, proper initial conditions and the time development.
\item There exists a (covariant) lower bound of energy for tachyons; in 
terms of the contravariant zero-component of the four-momentum this lower 
bound is simply zero.
\item The paradox of ``transcendental'' tachyons, apparent in the standard 
approach, disappears.
\item Tachyonic field can be consistently quantized using the CT 
synchronization scheme.
\item Tachyons distinguish a preferred frame {\em via\/} mechanism of the 
spontaneous symmetry breaking \cite{Rem:tac,Rem2}; consequently the 
relativity principle is broken, but the Lorentz covariance (and symmetry) is 
preserved. The preferred frame can be identified with the cosmic background 
radiation frame.
\item Classification of all possible unitary Poincar\'e mappings for 
space-like momenta is given. The important and unexpected result is that 
unitary orbits for space-like momenta are induced from the $SO(2)$ little 
group. This holds because we have a bundle of Hilbert spaces rather than a 
single Hilbert space of states. Therefore unitary operators representing 
Poincar\'e group act in irreducible orbits in this bundle. Each orbit is 
generated from subspace with $SO(2)$ stability group. Consequently, 
elementary states are labelled by helicity, in an analogy with the 
light-like case. This fact is extremely important because we have no problem 
with infinite component fields.
\item A consistent quantum field theory for tachyons with helicity $\lambda 
= 0$ and $\lambda=\pm\frac{1}{2}$ is formulated.
\end{itemize}
In the paper \cite{CibRem} the $\beta$-decay amplitude is calculated under 
assumption that neutrino is a tachyon.

\section{Preliminaries}\label{pre}
As is well known, in the standard framework of the special relativity, 
space-like geodesics do not have their physical counterparts. This is an 
immediate consequence of the assumed causality principle which admits 
time-like and light-like trajectories only.

In the papers by Terletsky \cite{Ter}, Tanaka \cite{Tnk}, Sudarshan 
\emph{et al}.\ \cite{BDS}, Recami \emph{et al}.\ \cite{Rec1,OR,Rec2} and 
Feinberg \cite{Fei} the causality problem has been reexamined and a 
physical interpretation of space-like trajectories was introduced. 
However, every proposed solution raised new unanswered questions of the 
physical or mathematical nature \cite{T&M}. The difficulties are specially 
frustrating on the quantum level \cite{KK,KK2,Nak}. 

It is rather evident that a consistent description of tachyons lies in a 
proper extension of the causality principle. Notice that interpretation of 
the space-like world lines as physically admissible tachyonic trajectories 
favour the  constant-time initial hyperplanes. This follows from the fact 
that only such surfaces intersect each world line with locally nonvanishing 
slope once and only once. Unfortunately, the instant-time hyperplane is 
\emph{not a Lorentz-covariant notion} in the standard formalism, which is 
just the source of many troubles with causality. 

The first step toward a solution of this problem can be found in the 
papers by Chang \cite{Cha1,Cha2,Cha3}, who introduced four-dimensional 
version of the Tangherlini transformations \cite{Tan}, termed the 
Generalized Galilean Transformations (GGT). In \cite{Rem1} it was shown 
that GGT, extended to form a group, are hidden (nonlinear) form of the 
Lorentz group transformations with $SO(3)$ as a stability subgroup. 
Moreover, a difference with the standard formalism lies in a nonstandard 
choice of the synchronization procedure. As a consequence a constant-time 
hyperplane is a covariant notion. In the following we will call this  
procedure of synchronization the \emph{Chang--Tangherlini synchronization 
scheme}. 

It is important to stress the following two well known facts: (a)~the 
definition of a coordinate time depends on the synchronization scheme 
\cite{Rei,TV,Var}, (b)~synchronization scheme is a convention, because no 
experimental procedure exists which makes it possible to determine the 
one-way velocity of light without use of superluminal signals \cite{Jam}. 
Notice that a choice of a synchronization scheme, different that the 
standard one, \emph{does not affect seriously the assumptions of special 
relativity but evidently it can change the causality notion}, depending on 
the definition of the coordinate time.

As it is well known, intrasystemic synchronization of clocks in their 
``setting'' (zero) requires a definitional or conventional 
stipulation---for discussion see Jammer \cite{Jam}, Sj\"odin \cite{Sjo} 
(see also \cite{MS}). Really, to determine one-way light speed it is 
necessary to use synchronized clocks (at rest) in their ``setting'' 
(zero)\footnote{Evidently, without knowledge of the one-way light speed, 
it is possible to synchronize clocks in their rate only \cite{And}.}. On 
the other hand to synchronize clocks we should know the one-way light 
velocity. Thus we have a logical loophole. In other words no experimental 
procedure exists (if we exclude superluminal signals) which makes possible 
to determine unambiguously and without any convention the one-way velocity 
of light (for analysis of some experiments see Will \cite{Wil1}). 
Consequently, an \emph{operational meaning has the average value of the 
light velocity around closed paths only}. This statement is known as the 
conventionality thesis \cite{Jam}. Following Reichenbach \cite{Rei}, two 
clocks $\sf A$ and $\sf B$ stationary in the points $A$ and $B$ of an 
inertial frame are defined as being synchronous with help of light signals 
if $t_B=t_A+\varepsilon_{AB}(t'_A-t_A)$. Here $t_A$ is the emission time of 
light signal at point $A$ as measured by clock $\sf A$, $t_B$ is the 
reception-reflection time at point $B$ as measured by clock $\sf B$ and 
$t'_A$ is the reception time of this light signal at point $A$ as measured 
by clock $\sf A$. The so called synchronization coefficient 
$\varepsilon_{AB}$ is an arbitrary number from the open interval $(0,1)$. 
In principle it can vary from point to point. The only conditions for 
$\varepsilon_{AB}$ follow from the requirements of symmetry and 
transitivity of the synchronization relation. Note that 
$\varepsilon_{AB}=1-\varepsilon_{BA}$. The one-way velocities of light 
from $A$ to $B$ ($c_{AB}$) and from $B$ to $A$ ($c_{BA}$) are given by 
\begin{displaymath}
c_{AB}=\frac{c}{2\varepsilon_{AB}},\quad
c_{BA}=\frac{c}{2\varepsilon_{BA}}.
\end{displaymath}
Here $c$ is the round-trip average value of the light velocity. In 
standard  synchronization $\varepsilon_{AB}=\frac{1}{2}$ and consequently 
$c=c_{AB}$ for each pair $A$, $B$. 

The conventionality thesis states that from the operational point of view 
the choice of a fixed set of the coefficients $\varepsilon$ is a 
convention. However, the explicit form of the Lorentz transformations will 
be $\varepsilon$-dependent in general. The question arises: Are equivalent 
notions of causality connected with different synchronization schemes? As 
we shall see throughout this work the answer is \emph{negative} if we 
admit tachyonic world lines. In other words, the causality requirement,  
logically independent of the requirement of the Lorentz covariance, can  
contradict the conventionality thesis and consequently it can prefer a  
definite synchronization scheme, namely CT scheme if an absolute causality 
is assumed.

It is very interesting, that in the framework of CT synchronization two 
oldstanding theoretical problems \cite{Bac}, completely unconnected with 
tachyons, have solutions; namely: (a)~the manifestly covariant canonical 
formalism for relativistic particle can be found and (b)~it is possible to 
construct a covariant position operator and set of covariant relativistic 
localizable states \cite{CR}.

\section{The Chang--Tangherlini synchronization}
As was mentioned in Section \ref{pre}, in the paper by Tangherlini \cite{Tan} 
a family of inertial frames in $1+1$ dimensional space of events was 
introduced with the help of transformations which connect the time 
coordinates by a simple (velocity dependent) rescaling. This construction was 
generalized to the $1+3$ dimensions by Chang \cite{Cha1,Cha2}. As was shown 
in the paper \cite{Rem1}, the Chang--Tangherlini inertial frames can be 
related by a group of transformations isomorphic to the orthochronous Lorentz 
group. Moreover, the coordinate transformations should be supplemented by 
transformations of a vector-parameter interpreted as the velocity of a 
privileged frame. It was also shown that the above family of frames is 
equivalent to the Einstein--Lorentz one; (in a contrast to the interpretation 
in \cite{Cha1,Cha2}). A difference lies in another synchronization procedure 
for clocks \cite{Rem1}. In the Appendix we derive realization of the Lorentz 
group given in \cite{Rem1} in a systematic way \cite{Rem2}.

Let us start with a simple observation that the description of a family 
of inertial frames in the Minkowski space-time is not so natural. Instead, it 
is obvious that the geometrical notion of bundle of frames is more natural. 
Base space is identified with the space of velocities; each velocity marks 
out a coordinate frame. Indeed, from the point of view of an observer (in a 
fixed inertial frame) all inertial frames are labelled by their velocities 
with respect to him. Therefore, in principle, to define the transformation 
rules between frames, we should use, except of coordinates, also this 
vector-parameter, related to velocities of frames with respect to a 
distinguished observer. 

Notice that a distinguishing of a preferred inertial frame is in full 
agreement with \emph{local} properties of the observed expanding world. 
Indeed, we can fix a local frame in which the Universe appears spherically; 
it can be done, in principle, by investigation of the isotropy of the Hubble 
constant \cite{Wei}. It concides with the cosmic background radiation frame. 
Thus it is natural to ask for a formalism incorporating locally Lorentz 
symmetry and the existence of a preferred frame.\footnote{Frequently, the 
notion of preferred frame is associated with a violation of Lorentz 
invariance \cite{Wil2} but in our case the Lorentz invariance is assumed to 
be exact.}

Below we list our basic requirements:
\begin{enumerate}
\item Coordinate frames are related by a set of transformations isomorphic to 
the Lorentz group (\textbf{Lorentz covariance}).\label{i}
\item The average value of the light speed over closed paths is constant 
($c$) for all inertial observers (\textbf{constancy of the round-trip light 
velocity}).\label{ii}
\item With respect to the rotations $x^0$ and $\vec{x}$ transform as $SO(3)$ 
singlet and triplet respectively (\textbf{isotropy}).\label{iii}
\item Transformations are linear with respect to the coordinates 
(\textbf{affinity}).\label{iv}
\item We admit an additional set of parameters $u$ (\textbf{velocity 
space---the base space} for the bundle of inertial frames).\label{5}
\end{enumerate}
We see that assumptions \ref{i}--\ref{iv} are the standard ones. In the 
following we consider also two distinguished cases corresponding to the 
relativity principle and absolute causality requirements respectively. 
Hereafter we shall use the natural units $c = \hbar = 1$.

\subsection{Poincar\'e group transformation rules in the CT synchronization}
According to our assumptions, transformations between two coordinate frames 
$x^{\mu}$ and ${x'}^{\mu}$ have the following form
\begin{equation}\label{x'}
x'(u') = D(\Lambda, u) (x(u) + a).
\end{equation}
Here $D(\Lambda, u)$ is a real (invertible) $4\times4$ matrix, $\Lambda$ 
belongs to the Lorentz group and $u^{\mu}$ is assumed to be four-velocity of 
distinguished frame, i.e., it transforms like $dx^{\mu}$
\begin{equation}\label{u'}
u' = D(\Lambda, u) u.
\end{equation}
The $a^\mu$ are translations in the frame $x^\mu(u)$.
It is easy to verify that the transformations (\ref{x'})--(\ref{u'}) 
constitute a realization of the Lorentz group if the following composition 
law holds
\begin{equation}
D\left(\Lambda_2, D(\Lambda_1, u) u\right)D(\Lambda_1, u) = D(\Lambda_2 
\Lambda_1, u).
\end{equation}
Now, because $u^{\mu}$ is assumed to be four-velocity of an inertial frame, 
it must be related to a time-like Lorentzian four-velocity\footnote{Both $u$ 
and $u_E$ must be related to the quotient space $SO(3,1)/SO(3)$ --- the base 
space of the frame bundle under consideration.} $u_E$ ($u^2_E = 1$); 
subscript $E$ means Einstein--Poincar\'e synchronization\footnote{In the 
papers by Chang \cite{Cha1,Cha2,Cha3} it
was used some kinematical objects with an unproper physical interpretation 
\cite{SP,FN}. For this reason we should be precise in the nomenclature 
related to different synchronizations.} (EP synchronization), where $u_E$ 
has the standard transformation law
\begin{equation}\label{uE}
u_E'=\Lambda u_E.
\end{equation}
Let us denote the intertwining matrix by $T(u)$, i.e. 
\begin{equation}\label{uE:u}
u_E = T^{-1}(u) u.
\end{equation}

Therefore the explicit form of $D(\Lambda, u)$ satisfying the assumptions
\ref{i}--\ref{5} (see also the Appendix) is 
\begin{equation}\label{D}
D(\Lambda, u) = T(u') \Lambda T^{-1}(u),
\end{equation}
where
\begin{equation}\label{T}
T(u) = \left(\begin{array}{c|c} 1 & b(u) \vec{u}^{\mathrm{T}} \\[1ex]
\hline 0 & I \end{array}\right).
\end{equation}
Here $b(u)$ is rotationaly invariant function of $u$; the superscript 
$^{\mathrm{T}}$ denotes transposition. Furthermore, the transformed 
four-velocity $u'$ is determined from the relation (\ref{uE:u}) and the 
transformation law (\ref{uE}). Thus the square of the line element
\begin{equation}\label{ds2}
ds^2 = g_{\mu\nu}(u)\, dx^{\mu}\, dx^{\nu}
\end{equation}
with
\begin{equation}\label{g}
g(u) = \left(T(u) \eta T^{\mathrm{T}}(u)\right)^{-1},
\end{equation}
where the Minkowski tensor $\eta = \diag (+,-,-,-)$, is invariant under the 
transformations (\ref{x'})--(\ref{u'}). Now, by means of (\ref{ds2}) for null 
geodesics, it is easy to calculate the light velocity. To do this let us 
notice that
\begin{equation}
u^2 = g_{\mu\nu}(u) u^{\mu} u^{\nu} = 1.
\end{equation}
The velocity of light propagation in a direction $\vec{n}$ ($\vec{n}^2 = 1$) 
reads
\begin{equation}
\vec{c} = \frac{\vec{n}}{1 + \vec{n} \vec{u} b(u)},
\end{equation}
so the Reichenbach synchronization coefficient takes the form
\begin{equation}\label{eps}
\varepsilon(\vec{n}, \vec{u}) = \frac{1}{2} \left(1 + \vec{n} \vec{u} 
b(u)\right).
\end{equation}

Therefore the function $b(u)$ distinguishes between different 
synchronizations. The most interesting choices of $b(u)$ correspond to the 
Einstein--Poincar\'e synchronization and to the Chang--Tangherlini one.

In the first case (EP), $b(u) = 0$, i.e.\ $T(u) = I$, so $g(u) = \eta$, 
$\vec{c} = \vec{c}_E = \vec{n}$ and $x = x_E$, $u = u_E$ with the standard 
transformation law (\ref{uE}). 

Now, the CT case is obtained under condition that \emph{the instant-time 
hyperplane $x^0 = \mathrm{constant}$ is an invariant notion}, i.e.\ that 
${x'}^0 = {D(\Lambda, u)^0}_0 x^0$ so ${D(\Lambda, u)^0}_k = 0$. Thus from 
eqs.\ (\ref{D}), (\ref{T}) we have in this case
\begin{equation}\label{b}
b(u) = -u^0.
\end{equation}
In the following we assume the CT synchronization defined by eq.\ (\ref{b}). 
In this case
\begin{equation}\label{T:u}
T(u) = \left(\begin{array}{c|c} 1 & -\vec{u}^{\mathrm{T}} u^0 \\[1ex] \hline 
0 & I \end{array}\right).
\end{equation}
Thus the interrelation between coordinates in EP and CT synchronizations is 
given by
\begin{equation}\label{EP:CT}
x^0_E = x^0 + u^0 \vec{u} \vec{x}, \quad \vec{x}_E = \vec{x},
\end{equation}
i.e. the essential difference is in the coordinate time definition.

Now, by means of (\ref{uE}) and (\ref{T:u}), we have determined the form of 
the matrix $D(\Lambda, u)$ (i.e. the transformation law 
(\ref{x'})--(\ref{u'})); namely
\begin{description}\boldmath
\item[for rotations $R \in SO(3) \subset SO(3,1)$]\unboldmath\hfill
\begin{equation}\label{D:Ru}
D(R, u) = \left(\begin{array}{c|c} 1 & 0 \\ \hline 0 & R \end{array}\right);
\end{equation}
\item[for boosts]\hfill
\begin{equation}\label{D:Wu}
D(W, u) = \left(\begin{array}{c|c} \frac{1}{W^0} & 0 \\ \hline
-\vec{W} & I +\frac{\vec{W} \otimes \vec{W}^{\mathrm{T}}}%
{\left(1 + \sqrt{1 + (\vec{W})^2}\right)}
-\vec{W} \otimes \vec{u}^{\mathrm{T}} u^0 \end{array}\right).
\end{equation}
\end{description} 
Here $W^{\mu}$ is the four-velocity of the primed frame ($x'$) with respect 
to the initial one ($x$)\footnote{I.e. $W^{\mu} = \frac{dX^{\mu}}{ds}$, where 
$d\vec{X}$ is the displacement of the origin of the primed frame in the time 
$dX^0$ and can be expressed by $u$ and $u' = D(W, u) u$ as follows
\begin{equation*}
W^0 = \frac{u^0}{{u'}^0}, \qquad
\vec{W} = \frac{(u^0 + {u'}^0) (\vec{u} - \vec{u}')}{\left[1 + u^0 {u'}^0 (1 
+ \vec{u} \vec{u}')\right]}.
\end{equation*}}. We see that absolute causality can be introduced in our 
framework because the coordinate time is rescaled only by the positive factor 
$\frac{1}{W^0}$, i.e.
\begin{equation}\label{x'0}
{x'}^0 = \frac{1}{W^0} x^0.
\end{equation}
Moreover, taking $W^\mu = u^\mu$, we can check from (\ref{D:Wu}) that $u^\mu$ 
is the four-velocity of a distinguished (provileged) inertial frame (fixed by 
$\tilde{u} = (1, 0, 0, 0)$) as seen from the unprimed frame $(x)$.

Now, the covariant metric tensor $g(u)$ (\ref{g}) takes the form
\begin{equation}\label{g:u}
\left[g_{\mu\nu}(u)\right] = \left(\begin{array}{c|c} 1 & u^{0} 
\vec{u}^{\mathrm{T}} \\[1ex] \hline u^{0} \vec{u} & 
-I + \vec{u} \otimes \vec{u}^{\mathrm{T}} (u^{0})^{2} \end{array}\right),
\end{equation}
while the contravariant one reads
\begin{equation}
g^{-1}(u) = \left(\begin{array}{c|c}  (u^{0})^{2} & u^{0} 
\vec{u}^{\mathrm{T}} \\[1ex] \hline
u^{0} \vec{u} & -I \end{array}\right),
\end{equation}
so the square of the length has the Euclidian form $dl^2 = -g^{ik}\, dx^i\, 
dx^k = d\vec{x}^2$. Furthermore, the light velocity $\vec{c}$ is given by
\begin{equation}
\vec{c} = \frac{\vec{n}}{1 - \vec{n} \vec{u} u^0}
\end{equation}
and we can check that $\langle|\vec{c}|\rangle_{\text{closed path}} = 1$.

Now, by means of relations between differentials (see \eqref{EP:CT})
\begin{equation}\label{dx}
dx^0_E = dx^0 + u^0 \vec{u}\, d\vec{x},\qquad
d\vec{x}_E = d\vec{x},
\end{equation}
we obtain interrelations between velocities in both synchronizations; namely
\begin{gather}\label{v:vE}
\vec{v} = \frac{\vec{v}_E}{1 - \frac{\vec{v}_E \vec{u}_E}{u^0_E}},
\\ \label{vE:v}
\vec{v}_E = \frac{\vec v}{1 + \vec{v} \vec{u} u^0}.
\end{gather}
Notice, that for $|\vec{v}| > |\vec{c}|$ the above formulas have a 
singularity.

We can also express the transformation matrix (\ref{D:Wu}) by means of the 
velocities $\vec{\sigma} = \frac{\vec{u}}{u^0}$ and $\vec{V} = 
\frac{\vec{W}}{W^0}$ via the relation
\begin{equation}\label{gam:v}
\frac{1}{W^0} = \sqrt{\left(1 + \vec{\sigma} \vec{V} \gamma^{-2}_0\right)^2 - 
(\vec{V})^{2}}
\end{equation}
with
\begin{equation}\label{gam0}
\gamma_0 = \left[\frac{1}{2} \left(1 + \sqrt{1 + (2
\vec{\sigma})^2}\right)\right]^{1/2}.
\end{equation}

Finally, let us notice that a second rank tensor, say ${\theta^\mu}_\nu$, 
transforms under the transformation law (\ref{x'})--(\ref{u'}) according to
\begin{equation}\label{th'}
\theta'(u') = D(\Lambda, u) \theta(u) D^{-1}(\Lambda, u).
\end{equation}
Therefore, taking into account the triangular form (\ref{D:Ru})--(\ref{D:Wu}) 
of $D(\Lambda, u)$, it is easy to see that the following conditions are 
invariant under (\ref{th'})
\begin{enumerate}
\item ${\theta^0}_k = 0$ (this implies ${\theta^0}_0 = \mathrm{const}$);
\item ${\theta^0}_k = 0$, ${\theta^0}_0 = \mathrm{const}$, ${\theta^i}_j = 
\alpha {\delta^i}_j$, where $\alpha$ is a scalar function.
\end{enumerate}
The second condition in the form ${\theta^0}_\mu = 0$, ${\theta^i}_j = 
-{\delta^i}_j$ is crucial for construction of a \emph{covariant} canonical 
formalism for tachyons. This will be done in the Sec.~\ref{can}.

\subsection{Causality and kinematics in the CT synchronization}\label{caus}
In CT scheme \emph{causality has an absolute meaning}. This follows from
the transformation law (\ref{x'0}) for the coordinate time:
$x^0$ is rescaled by a positive, velocity dependent, factor $\frac{1}{W^0}$.
Thus this formalism extends the EP causality by allowing faster than
light propagation. It can be made transparent if we consider the
relation derived from eq.\ (\ref{dx})
\begin{equation}\label{D19}
\frac{dx^0}{dx^0_E} = 1 - \frac{\vec{u}_E \vec v_E}{u^0_E}.
\end{equation}
For $|\vec v_E| \leq 1$ we have $\frac{dx^0}{dx^0_E} > 0$, so the EP and CT 
causality coincide in this case. Nevertheless for $|\vec v_E| > 1$, the sign 
of $\frac{dx^0}{dx^0_E}$ is indefinite which is a consequence of an 
inadequacy of the EP synchronization to description of faster than light 
propagation. On the other hand in the CT synchrinization framework, by means 
of the eq.\ (\ref{x'0}), each time interval $\Delta{x^0}$ is 
observer-independent, so there are no causal problems for tachyons.

Let us consider in detail a free particle case associated with a space-like 
geodesics. The corresponding action $S$ is of the form
\begin{equation}\label{S}
S_{12} = -\kappa \int^{\lambda_2}_{\lambda_1} \sqrt{-ds^2}
\end{equation}
where the square of the space-like line element
\begin{equation}\label{ds2<0}
ds^2 = g_{\mu\nu}(u) \frac{dx^\mu}{d\lambda} \frac{dx^\nu}{d\lambda} 
d\lambda^2 < 0
\end{equation}
and the continuous affine paramenter $\lambda$ is defined along the 
trajectory as monotonically increasing as one proceeds along the curve in a 
fixed direction.

The equations od motion are obtained by means of the variational principle 
and reads
\begin{equation}\label{d/dl}
\frac{d}{d\lambda} \left(\frac{\dot{x}^\mu}{\sqrt{-g_{\mu\nu}(u) \dot{x}^\mu 
\dot{x}^\nu}}\right) = 0 
\end{equation}
with $\dot{x}^\mu = \frac{dx^\mu}{d\lambda} \equiv w^\mu$. Now, we are free 
to take the path parameter as $d\lambda = \sqrt{-ds^2}$, so the four-velocity 
$w^\mu$ satisfies
\begin{equation}\label{w2}
w^2 = g_{\mu\nu}(u) w^\mu(u) w^\nu(u) = -1
\end{equation}
and consequently
\begin{equation}\label{w.}
\dot{w}^\mu = \ddot{x}^\mu = 0.
\end{equation}

Let us focus our attention on the constraint (\ref{w2}). Obviously it 
defines an one-sheet hyperboloid; in particular in the preferred frame (for 
$u = \tilde{u} = (1, 0, 0, 0)$) $g_{\mu\nu}(\tilde{u}) = \eta_{\mu\nu}$, so 
$\eta_{\mu\nu} w^\mu(\tilde{u}) w^\nu(\tilde{u}) = -1$, like in the Einstein 
synchronization. However, there is an important difference; namely under 
Lorentz boosts the zeroth component $w^0(u)$ of $w^\mu$ is rescaled by a 
positive factor only (see eq.\ (\ref{x'0})) i.e. ${w'}^0(u') = \frac{1}{W^0} 
w^0(u)$. Therefore, contrary to the Einstein--Poincar\'e synchronization, in 
this case points of the upper part of the hyperboloid (\ref{w2}) (satisfying 
$w^0(u) > 0$) transforms again into points of the upper part. This allows us 
to define consistently the velocity of a tachyon:
\begin{equation}\label{v}
\vec{v} = \frac{d\vec{x}}{dx^0} = \frac{\vec{w}}{w^0}
\end{equation}
because now, for each observer, the tachyon speed is finite (i.e. $|\vec{v}| 
< \infty$, $w^0 > 0$). We see that the infinite velocity is a limiting 
velocity, like in the non-relativistic case (it corresponds to $w^0 = 0$ 
which is an invariant condition). Notice that the constraint relation 
(\ref{w2}) implies that velocity of a tachyon moving in a direction $\vec{n}$ 
is restricted by the inequality
\begin{equation}
|\vec{c}| = \frac{1}{1 - \vec{n} \vec{u} u^0} < |\vec{v}| < \infty.
\end{equation}
Furthermore, the transformation law for velocities in the EP synchronization, 
derived from (\ref{x'}), (\ref{u'}), (\ref{D:Wu}) reads
\begin{equation}\label{v'}
\vec{v}' = W^0 \left[\vec{v} + \vec{W} \left(\frac{(\vec{W} \vec{v})}{\left(1 
+ \sqrt{1 + (\vec{W})^2}\right)} - u^0 (\vec{u} \vec{v}) - 1\right)\right]
\end{equation}
We see that the transformation law (\ref{v'}) is well defined for all 
velocities (sub- and superluminal). Recall that in the EP synchronization the 
corresponding transformation rule reads
\begin{equation}\label{vE'}
\vec{v}'_E = \frac{\vec{v}_E - \vec{W}_E \left(1 - \frac{\vec{W}_E 
\vec{v}_E}{(1 + W^0_E)}\right)}{W^0_E - \vec{W}_E \vec{v}_E}
\end{equation}

We observe that the denominator in the first part of the above transformation 
rule can vanish for $|\vec{v}_E| > 1$. Thus a tachyon moving with $1 < 
|\vec{v}_E| < \infty$ can be converted by a finite Lorentz map into a 
``transcendental'' tachyon with $|\vec{v}_E'| = \infty$. This discontinuity 
is an apparent inconsistency of this transformation law; namely in the EP 
scheme \emph{tachyonic velocity space does not constitute a representation 
space for the Lorentz group}. A technical point is that the space-like 
four-velocity cannot be related to a three-velocity in this case by the 
relation $\vec{v}_E = \frac{\vec{w}_E}{w^0_E}$, because  $w^0_E$ can take 
the value zero for a finite Lorentz transformation.

Concluding, in the CT synchronization the problem of ``transcendental'' 
ta\-chyons does not appear---contrary to the eq.\ (\ref{vE'}), the 
transformation law (\ref{v'}) is continuous, does not ``produce'' 
``transcendental'' tachyons and completed by rotations, forms (together with 
the mapping  $u \to u'$) a realization of the Lorentz group and the relation 
of $\vec{v}$ to the four-velocity is nonsingular. 

\subsection{Canonical formalism}\label{can}
Let us identify the Lagrangian of a free tachyon related to the action 
(\ref{S}); by means of the formulas (\ref{g:u}), (\ref{ds2}), (\ref{v}) we 
have
\begin{equation}
L = \kappa \sqrt{(\vec{v})^2 - (1 + u^0 \vec{u} \vec{v})^2}
\end{equation}
Thus the canonical momenta read
\begin{equation}
\pi_k = \frac{\partial L}{\partial v^k} = \frac{\kappa \left[v^k - u^k u^0 (1 
+ u^0 \vec{u} \vec{v})\right]}{\sqrt{(\vec{v})^2 - (1 + u^0 \vec{u} 
\vec{v})^2}} = -\kappa \omega_k
\end{equation}
where we have used eq.\ \eqref{w2}.
The Hamiltonian
\begin{equation}
H = \pi_k v^k - L = \frac{\kappa (1 + u^0 \vec{u} \vec{v})}{\sqrt{(\vec{v})^2 
- (1 + u^0 \vec{u} \vec{v})^2}} = +\kappa \omega_0
\end{equation}
Therefore the covariant four-momentum $k_\mu$ of tachyon can be defined as
\begin{equation}\label{k0:om}
k_0 = H = \kappa \omega_0, \quad
\nad{k} = -\nad{\pi} = \kappa \nad{\omega}
\end{equation}
i.e. $k_\mu = \kappa \omega_\mu$.

Notice that
\begin{equation}\label{k2}
k^2 = g^{\mu\nu}(u) k_\mu k_\nu = -\kappa^2
\end{equation}
and the energy $H = \kappa \omega_0$ has in each inertial frame a finite 
lower bound corresponding to $|\vec{v}| \to \infty$, i.e.
\begin{equation}
E > \frac{\kappa \sqrt{1 - (u^0)^2} \cos\phi}{\sqrt{1 - \left(\sqrt{1 - 
(u^0)^2} \cos\phi\right)^2}} \equiv \mathcal{E}(u^0, \phi)
\end{equation}
where $\cos\phi = \frac{\vec{u} \vec{v}}{|\vec{u}| |\vec{v}|}$.

Therefore, contrary to the standard case, the energy of tachyon is always 
restricted from below by $\mathcal{E}(u^0, \phi) > -\infty$. Moreover, if we 
calculate the contravariant four-momentum $k^\mu = g^{\mu\nu}(u) k_\nu = 
\kappa \omega^\mu$ we obtain that
\begin{equation}
k^0 = \frac{\kappa}{\sqrt{(\vec{v})^2 - (1 + u^0 \vec{u} \vec{v})^2}} > 0
\end{equation}
which confirm our statement that the sign of $k^0$ is Lorentz invariant also 
for tachyons (recall the transformation law (\ref{x'0})). In terms of 
$\nad{k}$ the zeroth components of $k$ read
\begin{gather}\label{k0u}
k^0 = u^0 \sqrt{(\vec{u} \nad{k})^2 + (|\nad{k}|^2 - \kappa^2)},\\ 
\label{k0d}
k_0 = \frac{1}{u^0} \left(-\vec{u} \nad{k} + \sqrt{(\vec{u} \nad{k})^2 + 
(|\nad{k}|^2 - \kappa^2)}\right)
\end{gather}
and the range of the covariant momentum $\nad{k}$ is determined by the 
inequality
\begin{equation}\label{range}
|\nad{k}| > \kappa \left(1 + \left(\frac{\vec{u} 
\nad{k}}{|\nad{k}|}\right)^2\right)^{-\frac{1}{2}}
\end{equation}
i.e. values of $|\nad{k}|$ lie outside of the oblate spheroid with half-axes 
$\kappa$ and $\kappa u^0$ and with the symmmetry axis parallel to $\vec{u}$.

Now, the Hamilton equations have the form
\begin{align}\label{H1}
\frac{d\vec{x}}{dt} &= \frac{\partial H}{\partial\nad{\pi}} = -\frac{\partial 
k_0}{\partial\nad{k}} = \frac{\vec{k}}{k^0} = \vec{v}, \\[1ex] \label{H2}
\frac{d\nad{k}}{dt} &= -\frac{\partial H}{\partial\vec{x}} = 0.
\end{align}
From the second equation it follows that $\frac{d\vec{v}}{dt} = 0$.

Furthermore, the most general Poincar\'e covariant Poisson bracket reads
\begin{equation}\label{Pb}
\{A, B\} = {C^\mu}_\nu \left(\frac{\partial A}{\partial x^\mu} \frac{\partial 
B}{\partial k_\nu} - \frac{\partial B}{\partial x^\mu} \frac{\partial 
A}{\partial k_\nu}\right)
\end{equation}
where ${C^\mu}_\nu$ is a second rank tensor. Because we assume parity and 
translational invariance, its most general form is
\begin{equation*}
{C^\mu}_\nu = a {\delta^\mu}_\nu + b k^\mu u_\nu + d u^\mu k_\nu + e k^\mu 
k_\nu + f u^\mu u_\nu
\end{equation*}
where $a$, $b$, $d$, $e$, $f$ are Lorentz scalars.

Notice that (\ref{Pb}) imply
\begin{equation*}
\{x^\mu, k_\nu\} = {C^\mu}_\nu.
\end{equation*}
However, we should demand
\begin{alignat}{2}\label{con1}
\{x^0, k_\nu\} &= 0 &\quad \text{i.e. } {C^0}_\nu &= 0, \\ \label{con2}
\{x^i, k_k\} &= -{\delta^i}_k & \text{i.e. } {C^i}_k &= -{\delta^i}_k, \\
\intertext{as well as} \label{con3}
\{x^i, k_0\} &= \frac{\vec{k}}{k^0} & \text{i.e. } {C^i}_0 &= 
\frac{\vec{k}}{k^0}.
\end{alignat}
The first condition tells that the coordinate time $t = x^0$ is not any 
dynamical variable, but a dynamical parameter. The second one is evidently 
the statement that $\vec{x}$ and $\nad{\pi} = -\nad{k}$ are canonically 
conjugated. Finally the last requirement follows from the Hamilton equations 
\eqref{H1}--\eqref{H2}. The above conditions determine the Poisson bracket 
\eqref{Pb} which takes the form
\begin{equation}\label{Pb2}
\{A, B\} = -\left({\delta^\mu}_\nu - \frac{k^\mu u_\nu}{u k}\right) 
\left(\frac{\partial A}{\partial x^\mu} \frac{\partial B}{\partial k_\nu} - 
\frac{\partial B}{\partial x^\mu} \frac{\partial A}{\partial k_\nu}\right)
\end{equation}
with $u k = u_\mu k^\nu = u_0 k^0$; this last equality follows from the fact 
that $u_k = g_{k\mu}(u) u^\mu = 0$.

It is easy to see that the Poisson bracket defined by the relation 
\eqref{Pb2} satisfies all necessary conditions:
\begin{itemize}
\item It is linear with respect to the both factors, antisymmetric, 
satisfying the Leibniz rule and fulfill the Jacobi identity.
\item It is manifestly Poincar\'e covariant in the CT synchronization 
(recall the comment after the eq.\ \eqref{th'}).
\item It is consistent with all canonical conditions 
\eqref{con1}--\eqref{con3}.
\item It is easy to check that the tachyonic dispersion relation \eqref{k2}, 
$k^2 = -\kappa^2$, is consistent with this bracket. i.e. $\{k^2, k_\nu\} = 
\{k^2, x^\mu\} = 0$; therefore we do not need to introduce a Dirac bracket.
\end{itemize}
It is clear that in analogous way we can construct canonical formalism for 
bradyons ($k^2 = m^2$) in the CT synchronization. We also are able to 
introduce in this scheme a covariant position operator on the quantum ground 
\cite{CR}.

\subsection{Synchronization group and the relativity principle}\label{syn}
From the foregoing discussion we see that the CT synchronization prefers a 
privileged frame corresponding to the value $\vec{u}=0$. It is clear that if 
we forget about tachyons such a preference is only formal; namely we can 
choose each inertial frame as a preferred one.

Let us consider two CT synchronization schemes, say $A$ and $B$, under two 
different choices of privileged inertial frames, say $\Sigma_A$ and 
$\Sigma_B$. Now, in each inertial frame $\Sigma$ two coordinate charts $x_A$ 
and $x_B$ can be introduced, according to both schemes $A$ and $B$ 
respectively. The interrelation is given by the almost obvious relations 
\begin{gather}\label{xB}
x_B = T(u^B) T^{-1}(u^A) x_A, \\
\label{uB}
u^B = D(\Lambda_{BA}, u^A) u^A,
\end{gather}
where $u^A$ ($u^B$) is the four-velocity of $\Sigma_A$ ($\Sigma_B$) with 
respect to $\Sigma$. $T(u)$ is given by the eq.\ (\ref{T:u}). We observe that 
a set of all possible four-velocities $u$ must be related by Lorentz group 
transformations too, i.e.\ $\{\Lambda_{BA}\}=L_S$. Of course it does not 
coincide with our intersystemic Lorentz group $L$. We call the group $L_S$ a
synchronization group \cite{Rem:tac,Rem2}.

Now, if we compose the transformations (\ref{x'}), (\ref{u'}) of $L$ and 
(\ref{xB})--(\ref{uB}) of $L_S$ we obtain
\begin{gather}\label{x'2}
x' = T(u') \Lambda T^{-1}(u) x,\\ \label{u'2}
u' = D(\Lambda_S\Lambda, u) u
\end{gather}
with $\Lambda_S \in L_S, \Lambda \in L$.

Therefore \eqref{x'2}--\eqref{u'2} can be obtained as a composition of two 
mutually commuting transformations
\begin{align}\label{L0}
L_0 \ni (I, \Lambda_0) &\colon \left\{\begin{array}{l} x' = T(u) \Lambda_0 
T^{-1}(u) x \\
u' = u \end{array}\right. \\ \label{LS}
L_S \ni (\Lambda_S, I) &\colon \left\{\begin{array}{l} x' = T(u') T^{-1}(u) x 
\\
u' = D(\Lambda_s, u) u \end{array}\right.
\end{align}

Thus the composition law for $(\Lambda_S, \Lambda_0)$ reads
\begin{equation}\label{LL}
(\tilde{\Lambda}_S, \tilde{\Lambda}_0)(\Lambda_S, \Lambda_0) = 
(\tilde{\Lambda}_S \Lambda_S, \tilde{\Lambda}_0 \Lambda_0).
\end{equation}
It is evident that both $L_S$ and $L_0$ are isomorphic to the Lorentz group. 
Therefore the set $\{(\Lambda_S,\Lambda)\}$ is the direct product of two 
Lorentz groups $L_0\otimes L_S$. The intersystemic Lorentz group $L$ is the 
diagonal subgroup in this direct product, i.e. elements of $L$ are of the 
form $(\Lambda, \Lambda)$. Thus $L$ acts as an authomorphism group of $L_S$.

Now, the synchronization group $L_S$ realizes in fact the relativity 
principle. In our language the relativity principle can be formulated as 
follows: \emph{Any inertial frame can be choosen as a preferred frame}. What 
happens, however, when the tachyons do exist? In that case the relativity 
principle is obviously broken: \emph{If tachyons exist then one and only one 
inertial frame \textbf{must be} a preferred frame} to  preserve an absolute 
causality. Moreover, the one-way light velocity becomes a real, measured 
physical quantity because conventionality thesis breaks down. It means that 
the synchronization group $L_S$ is broken to the $SO(3)_{u}$ subgroup 
(stability group of $u$); indeed, transformations from the $L_S/SO(3)_{u}$ do 
not leave the causality notion invariant. As we show later, on the quantum 
level we have to deal with \emph{spontaneous breaking} of $L_S$ to $SO(3)$.

Notice, that in the real world a preferred inertial frames are distinguished 
\emph{locally} as the frames related to the cosmic background radiation. Only 
in such frames the Hubble constant is direction-independent.

\section{Quantization}
The following two facts, true only in CT synchronization, are extremely 
important for quantization of tachyons:
\begin{itemize}
\item Invariance of the sign of the time component of the space-like 
four-mo\-men\-tum i.e. $\varepsilon(k^0)=\mbox{inv}$,
\item Existence of a covariant lower energy bound; in terms of the 
contravariant space-like four-momentum $k^{\mu}$, $k^{2}<0$, this lower bound 
is exactly zero, i.e . $k^{0}\geq0$ as in the time-like and light-like case.
\end{itemize}
This is the reason why an invariant Fock construction can be done in our case. 

\subsection{Tachyonic representations}
Our fundamental object will be a bundle of Hilbert spaces $\mathcal{H}_u$ 
corresponding to the bundle of the inertial frames. Here we classify unitary 
Poincar\'e mappings in this bundle of Hilbert spaces for a space-like 
four-momentum. Furthermore we find the corresponding canonical commutation 
relations. As result we obtain that tachyons correspond to unitary mappings 
which are induced from $SO(2)$ group rather than $SO(2,1)$ one. Of course, a 
classification of unitary orbits for time-like and light-like four-momentum 
is standard, i.e., it is the same as in EP synchronization; this holds 
because the relativity principle is working in these cases (synchronization 
group is unbroken).

As usually, we assume that a basis in a Hilbert space  $\mathcal{H}_u$
(fibre) of one-particle states consists of
the eigenvectors $\left|k,u;\dots\right>$ of the four-momentum
operators namely
\begin{equation}
P^{\mu} |k, u; \dots \rangle = k^{\mu} |k, u; \dots \rangle        
\end{equation}
where
\begin{equation}\label{k'k}
\langle k', u; \dots |k, u; \dots \rangle = 2 k^{0} \delta^{3}(\nad{k'} - 
\nad{k})
\end{equation}
i.e. we adopt a covariant normalization. The $k^{0} = g^{0\mu} k_{\mu}$ is 
positive and the energy $k_{0}$ is the corresponding solution of the 
dispersion relation \eqref{k2}; both $k^0$ and $k_0$ are given by 
\eqref{k0u}--\eqref{k0d}. In the following we will use the abbreviation 
$\omega_{\mathbf{k}} \equiv k^0$ also.

The covariant normalization in (\ref{k'k}) is possible because in CT 
synchronization the sign of $k^0$ is an invariant. Thus we have no problem 
with an indefinite norm in $\mathcal{H}_u$.

Now, $k u \equiv k_{\mu} u^{\mu}$ is an additional Poincar\'e invariant. 
Summarizing, irreducible family of unitary operators $U(\Lambda, a)$ in the 
bundle of Hilbert spaces $H_u$ acts on an orbit defined by the following 
covariant conditions
\begin{itemize}
\item $k^2 = -\kappa^2$;
\item $\varepsilon(k^0) = \text{inv}$, for physical representations $k^0 > 0$ 
so $\varepsilon(k^0) = 1$ which guarantee a covariant lower bound of energy;
\item $q \equiv u k = \text{inv}$, it is easy to see that $q$ is the energy 
of tachyon measured in the privileged frame.
\end{itemize}
As a consequence there exists an invariant, positive definite measure
\begin{equation}\label{dmu}
d\mu(k, \kappa, q) = d^4\!k\, \theta(k^0) \delta(k^2 + \kappa^2) \delta(q - u 
k).
\end{equation}

Let us return to the problem of classification of irreducible unitary 
mappings $U(\Lambda, a)$ from $\mathcal{H}_u$ to $\mathcal{H}_{u'}$:
\begin{equation*}
U(\Lambda, a) |k, u; \dots \rangle = |k', u'; \dots \rangle;
\end{equation*}
here the pair $(k, u)$ is transported along trajectories belonging to an 
orbit fixed by the above mentioned invariant conditions. To follow the 
familiar Wigner procedure of induction, one should find a stability group of 
the double $(k, u)$. To do this, let us transform $(k, u)$ to the preferred 
frame by the Lorentz boost $L_{u}^{-1}$. Next, in the privileged frame, we 
rotate the spatial part of the four-momentum to the $z$-axis by an 
appropriate rotation $R^{-1}_{\vec{n}}$. As a result, we obtain the pair 
$(k,u)$ transformed to the pair $(\tilde{k}, \tilde{u})$ with
\begin{equation}\label{ku}
\tilde{k} = \begin{pmatrix} q \\ 0 \\ 0 \\ \sqrt{\kappa^2 + q^2} 
\end{pmatrix}, \qquad \tilde{u} = \begin{pmatrix} 1 \\ 0 \\ 0 \\ 0 
\end{pmatrix}.
\end{equation}
It is easy to see that the stability group of $(\tilde{k}, \tilde{u})$ is the 
$SO(2) = SO(2,1) \cap SO(3)$ group. Thus tachyonic unitary representations 
should be induced from the $SO(2)$ instead of $SO(2,1)$ group! Recall that 
unitary representations of the $SO(2,1)$ non-compact group are infinite 
dimensional (except of the trivial one). As a consequence, local fields in 
the standard case are necessarily infinite component ones (except of the 
scalar one). On the other hand, in the CT synchronization case unitary 
representations for space-like four-momenta in our bundle of Hilbert spaces 
are induced from irreducible, one dimensional representations of $SO(2)$ in a 
close analogy with a light-like four-momentum case. They are labelled by 
helicity $\lambda$, by $\kappa$ and by $q$ ($\varepsilon(k^0) = 
\varepsilon(q)$ is determined by $q$; of course a physical choice is 
$\varepsilon(q) = 1$).

Now, by means of the familiar Wigner procedure we determine the Lorentz 
group action on the base vectors; namely
\begin{equation}
U(\Lambda) |k, u; \kappa, \lambda, q\rangle = e^{i \lambda \varphi(\Lambda, 
k, u)} |k', u'; \kappa, \lambda, q \rangle
\end{equation}
where
\begin{equation}
e^{i \lambda \varphi(\Lambda, k, u)} = U\left(R^{-1}_{\Omega \vec{n}} 
\Omega R_{\vec{n}}\right)
\end{equation}
with
\begin{equation}
\Omega = L^{-1}_{u'} \Lambda L_u.
\end{equation}
Here $k$ and $u$ transform according to the law \eqref{x'}, \eqref{u'}, 
\eqref{D:Ru}, \eqref{D:Wu}. The rotation $R_{\vec{n}}$ connects $\tilde{k}$ 
with $D(L^{-1}_{u}, u)k$, i.e.
\begin{equation}
R_{\vec{n}} \tilde{k} = D(L^{-1}_{u}, u) k = \bar{k}.
\end{equation}
Taking into account \eqref{ku} we can derive the explicit form of 
$R_{\vec{n}}$
\begin{equation}
R_{\vec{n}} = \begin{pmatrix} 1 & 0 & 0 & 0 \\ 0 & 1 + 
\frac{(n^1)^2}{n^3 - 1} & \frac{n^1 n^2}{n^3 - 1} & n^1 \\ 0 & 
\frac{n^1 n^2}{n^3 - 1} & 1 + \frac{(n^2)^2}{n^3 - 1} & n^2 \\ 0 & n^1 & n^2 
& n^3 \end{pmatrix}
\end{equation}
where $\vec{n} = \nad{\bar{k}}/|\nad{\bar{k}}|$. Notice that for rotations 
$S$ form the stability group $SO(2)$ $R_{S \vec{n}} = S R_{\vec{n}} S^{-1}$.

It is easy to check that $R^{-1}_{\Omega \vec{n}} \Omega R_{\vec{n}}$ is a 
Wigner-like rotation belonging to the stability group $SO(2)$ of $(\tilde{k}, 
\tilde{u})$ and determines the phase $\varphi$. By means of standard 
topological arguments $\lambda$ can take integer or half-integer values only 
i.e. $\lambda=0, \pm 1/2, \pm 1, \dots.$

Now, the orthogonality relation (\ref{k'k}) reads
\begin{equation}
\langle k', u; \kappa', \lambda', q'|k, u; \kappa, \lambda, q\rangle = 2 
\omega_{\mathbf{k}} \delta^{3} (\nad{k'} - \nad{k}) \delta_{\lambda', 
\lambda}. 
\end{equation}

\subsection{Canonical quantization}
Following the Fock procedure, we define canonical commutation relations
\begin{gather}\label{aa}
[a_{\lambda}(k, u), a_{\tau}(p, u)]_{\pm} = 
[a_{\lambda}^{\dagger}(k, u), a_{\tau}^{\dagger}(p, u)]_{\pm} = 0,
\\ \label{aa+}
[a_{\lambda}(k, u), a_{\tau}^{\dagger}(p, u)]_{\pm} = 
2 \omega_{\mathbf{k}} \delta(\nad{k} - \nad{p}) \delta_{\lambda\tau},
\end{gather}
where $-$ or $+$ means the commutator or anticommutator and corresponds
to the bosonic ($\lambda$ integer) or fermionic ($\lambda$ half-integer)
case respectively. Furthermore, we introduce a Poincar\'e invariant
vacuum $|0\rangle$ defined by
\begin{equation}\label{a0}
\langle 0|0\rangle = 1 \quad\text{and}\quad
a_{\lambda}(k, u) |0\rangle = 0.                              
\end{equation}
Therefore the one particle states
\begin{equation}
a_{\lambda}^{\dagger}(k, u) |0\rangle
\end{equation}
are the base vectors belonging to an orbit in our bundle of
Hilbert spaces iff
\begin{gather}\label{27a}
U(\Lambda) a_{\lambda}^{\dagger}(k, u) U(\Lambda^{-1}) = 
e^{i \lambda \varphi(\Lambda, k, u)} 
a_{\lambda}^{\dagger}(k', u'),
\\ \label{27b}
U(\Lambda) a_{\lambda}(k, u) U(\Lambda^{-1}) = 
e^{-i \lambda \varphi(\Lambda, k, u)} a_{\lambda}(k', u'),
\end{gather}
and
\begin{equation}\label{P:a}
[P_{\mu}, a_{\lambda}^{\dagger}(k, u)]_{-} = k_{\mu}^+ 
a_{\lambda}^{\dagger}(k, u).
\end{equation}
Notice that
\begin{equation}
P_{\mu} = \int d^4k\, \theta(k^0) \delta(k^2 + \kappa^2) k_{\mu} 
\left(\sum_{\lambda} a_{\lambda}^{\dagger}(k,u)a_{\lambda}(k,u)\right)   
\label{P}
\end{equation}
is a solution of (\ref{P:a}).

Let us determine the action of the discrete transformations, space and
time inversions, $P$ and $T$ and the charge conjugation $C$ on the
states $|k, u; \kappa, \lambda, q\rangle$.
\begin{align}
P |k, u; \kappa, \lambda, q\rangle &= \eta_{s} |k^{\pi}, u^{\pi}; \kappa, 
-\lambda, q\rangle,\label{Ps}\\
T |k, u; \kappa, \lambda, q\rangle &= \eta_{t} |k^{\pi}, u^{\pi}; \kappa, 
\lambda, q\rangle,\label{Ts}\\
C|k, u; \kappa, \lambda, q\rangle &= \eta_{c} |k, u; \kappa, \lambda, q
\rangle_{c},\label{Cs}
\end{align}
where $|\eta_{s}| = |\eta_{t}| = |\eta_{c}| = 1$, $k^{\pi} = (k^{0}, 
-\vec{k})$, $u^{\pi} = (u^{0}, -\vec{u})$, the subscript $c$ means the 
antiparticle state and $P$, $C$ are unitary, while $T$ is antiunitary.

Consequently the actions of $P$, $T$ and $C$ in the ring of the field
operators read
\begin{align}
P a^{\dagger}_{\lambda}(k, u) P^{-1} &= \eta_{s} 
a^{\dagger}_{-\lambda}(k^{\pi}, u^{\pi}), \label{Pa}\\
T a^{\dagger}_{\lambda}(k, u) T^{-1} &= \eta_{t} 
a^{\dagger}_{\lambda}(k^{\pi}, u^{\pi}), \label{Ta}\\
C a^{\dagger}_{\lambda}(k, u) C^{-1} &= \eta_{c} 
b^{\dagger}_{\lambda}(k^{\pi}, u^{\pi}), \label{Ca}
\end{align}
where $b_{\lambda} \equiv a^c_{\lambda}$---antiparticle operators.

Finally we can deduce also the form of the helicity operator:
\begin{equation}\label{lam}
\hat{\lambda}(u) = -\frac{\hat{W}^{\mu} u_{\mu}}{\sqrt{(P u)^2 - P^2}}
\end{equation}
where
\begin{equation*}
\hat{W}^{\mu} = \frac{1}{2} \varepsilon^{\mu\sigma\lambda\tau} 
J_{\sigma\lambda}
P_{\tau}
\end{equation*}
is the Pauli-Lubanski four-vector.

Notice that
\begin{align}
P \hat{\lambda}(u) P^{-1} &= -\hat{\lambda}(u^{\pi}),\label{Pl}\\
T \hat{\lambda}(u) T^{-1} &= \hat{\lambda}(u^{\pi}),\label{Tl}\\
C \hat{\lambda}(u) C^{-1} &= \hat{\lambda}(u),\label{Cl}
\end{align}
as well as
\begin{equation}\label{la}
[\hat\lambda(u), a^\dagger_{\lambda}(u, k)] = \lambda 
a^\dagger_{\lambda}(u, k).
\end{equation}
 
\subsection{Local fields}
As usually we define local tachyonic fields as covariant Fourier
transforms of the creation--annihilation operators. Namely
\begin{multline}\label{field}
\varphi_{\alpha}(x, u) = \frac{1}{(2\pi)^{\frac{3}{2}}}
\int_{0}^{\infty} dq \int d\mu(k, \kappa, q)\\
\sum_{\lambda}
\left[w_{\alpha\lambda}(k, u) e^{i k x} b_{\lambda}^{\dagger}(k, u) + 
v_{\alpha\lambda}(k, u) e^{-i k x} a_{\lambda}(k, u)\right],
\end{multline}
where the amplitudes $w_{\alpha\lambda}$ and $v_{\alpha\lambda}$ satisfy the 
set of corresponding consistency conditions (the Weinberg conditions). Here 
we sum irreducible Poincar\'e orbits labelled by selected helicities and over 
the invariant $q$. Thus the integration in (\ref{field}) reduces to the 
integration with the measure 
\begin{equation}\label{measure}
d\mu(k, \kappa) = d^{4}k\, \theta(k^{0}) \delta(k^{2} + \kappa^{2}) 
\end{equation}
i.e.\ to the  integration over the space of all initial conditions.

\section{Scalar tachyonic field and its plane-wave decomposition}\label{loc} 
Let us consider a hermitian, scalar field $\varphi(x, u)$ satysfying the 
Klein--Gordon equation with imaginary ``mass'' $i\kappa$, i.e.\
\begin{equation}\label{K-G}
\left(g^{\mu\nu}(u) \partial_{\mu} \partial_{\nu} - \kappa^2\right) 
\varphi(x, u) = 0. 
\end{equation}
The Fourier decomposition of the field $\varphi$ reads
\begin{equation}
\varphi(x, u) = \frac{1}{(2\pi)^{3/2}} \int d\mu(k, u) \left(e^{i k x}  
a^{\dagger}(k, u) + e^{-i k x} a(k, u)\right).
\end{equation}
Integrating with respect to $k_0$ we obtain
\begin{equation}\label{phi}
\varphi(x, u) = \frac{1}{(2\pi)^{3/2}} \int_{\Gamma} 
\frac{d^3\nad{k}}{2\omega_{\mathbf{k}}} \left(e^{i k x} a^{\dagger}(k, u) + 
e^{-i k x} a(k_{+}, u)\right)
\end{equation}
where the integration range $\Gamma$ is
determined by the eq.\ \eqref{range}.

The canonical commutation rules \eqref{aa}, \eqref{aa+} take the form
\begin{gather}\label{aa1}
\bigl[a(k, u), a(p, u)\bigr] = \bigl[a^{\dagger}(k, u), a^{\dagger}(p, 
u)\bigr] = 0,\\
\label{aa+1}
\bigl[a(k, u), a^{\dagger}(p, u)\bigr] = 2 \omega_{\mathbf{k}} \delta(\nad{k} 
- \nad{p}).
\end{gather}
By the standard procedure we obtain the commutation rule for $\varphi(x,u)$ 
valid for an arbitrary separation
\begin{equation}\label{comm}
\left[\varphi(x, u), \varphi(y, u)\right] = - i \Delta(x - y, u),
\end{equation}
where the analogon of the Schwinger function reads
\begin{equation}
\Delta(x,u) = \frac{- i}{(2\pi)^{3}} \int d^{4}k\, \delta(k^{2} + \kappa^{2}) 
\varepsilon(k^{0}) e^{ikx}.
\end{equation}
It is remarkable that $\Delta$ does not vanish for a space-like
separation which is a direct consequence of the faster-than-light
propagation of the tachyonic quanta. Moreover $\Delta(x,u)|_{x^{0}=0}=0$
and therefore no interference occurs between two measurements of
$\varphi$ at an instant time. This property is consistent with our
interpretation of instant-time hyperplanes as the initial ones.

Now, because of the absolute meaning of the arrow of time in the CT
synchronization we can introduce an invariant notion of the time-ordered
product of field operators. In particular the tachyonic propagator
 \[ {\Delta}_{F}(x-y,u)=-i\left<0\right|
T(\varphi(x,u)\,\varphi(y,u))\left|0\right> \]
 is given by
 \begin{equation}\label{29}
 {\Delta}_{F}(x,u)=-\theta(x^0)\,
{\Delta}^{-}(x,u)+\theta(-x^0)\,{\Delta}^{+}(x,u)
 \end{equation}
 with
 \begin{equation}\label{Q4}
{\Delta}^{\pm}(x,u)=\frac{\mp i}{(2\pi)^3} \int
d^4k\,\theta(\pm k^0)\delta(k^2+\kappa^2)e^{ikx}. 
 \end{equation}
 The above singular functions are well defined as distributions on the
space of ``well behaved'' solutions of the Klein--Gordon equation
(\ref{K-G}).  The role of the Dirac delta plays the generalized function
\begin{equation}
{\delta}^{4}_{\Gamma}(x - y) = \frac{1}{(2\pi)^3} \delta(x^0 - y^0) 
\int_{\Gamma} d^3\nad{k}\, e^{i \nad{k} (\vec{x} - \vec{y})}.
\end{equation}
The above form of ${\delta}^{4}_{\Gamma}(x)$ express impossibility of
the localization of tachyonic quanta. In fact, the tachyonic field does
not contain modes with momentum $\nad{k}$ inside the spheroid defined in
eq.\ (\ref{range}). Consequently, by the Heisenberg uncertainty relation,
an exact localization of tachyons is impossible.

Note also that
\begin{equation*}
\partial^0 \Delta(x - y, u) \delta(x^0 - y^0) = {\delta}^{4}_{\Gamma}(x - y)
\end{equation*}
so the equal-time canonical commutation relations for $\varphi(x, u)$ and its 
conjugate momentum $\pi(x, u) = \partial^0 \varphi(x, u)$ have the correct 
form
\begin{gather}\label{q25a}
\delta(x^0 - y^0) \left[\varphi(x, u), \varphi(y, u)\right] = 
\delta(x^0 - y^0) \left[\pi(x, u), \pi(y, u)\right] = 0,
\\ \label{q25b}
\left[\varphi(x, u), \pi(y, u)\right] \delta(x^0 - y^0) = 
i {\delta}^{4}_{\Gamma}(x - y)
\end{gather}
as the operator equations in the space of states.

To do the above quantization procedure mathematically more  precise, we
can use wave packets rather than the plane waves. Indeed, with a help of
the measure we can define the Hilbert space $\mathcal{H}_{u}$ of one particle 
states with the scalar product
\begin{equation}\label{C2}
(f, g)_{u} = \int d\mu(k,u)\, f^*(k, u)\, g(k, u) < \infty. 
\end{equation}
Now, using standard properties of the Dirac delta we deduce
\begin{equation}\label{XX3}
(f, g)_{u} = \int_{\Gamma} \frac{d^3\nad{k}}{2 \omega_{\mathbf{k}}} f^{*}(k, 
u) g(k, u).
\end{equation}
It is remarkable that for  $\omega_{\mathbf{k}} \to 0$ to preserve inequality 
$\|f\|^{2}_{u} < \infty$ the wave packets $f(k,u)$ rapidly decrease to zero. 
This means that probability of ``momentum localization'' of a tachyon in the 
infinite velocity limit is going to zero in agreement with our intuition. As 
usually we introduce the smeared operators
\begin{equation}\label{C3}
a(f, u) = (2 \pi)^{-3/2} \int d\mu(k, u)\, a(k, u) f^*(k, u) 
\end{equation}
and the conjugate ones. The canonical commutation rules 
\eqref{aa1}--\eqref{aa+1} take the form
\begin{gather}\label{C4a}
\bigl[a(f, u), a(g, u)\bigr] = \bigl[a^{\dagger}(f, u), a^{\dagger}(g, 
u)\bigr] = 0,
\\ \label{C4b}
\bigl[a(f, u), a^{\dagger}(g, u)\bigr] = (f, g)_{u}.
\end{gather}
We have also $a(f,u)\left|0\right> = 0$ and
$\left<f,u|g,u\right>=\left(f,g\right)_{u}$, where
$\left|f,u\right>=a^{\dagger}(f,u)\left|0\right>$. According to our 
assumption of scalarity of
$\varphi(x,u)$
 \begin{equation}\label{A}
 L\ni\Lambda:\;
\varphi'(x',u')=\varphi(x,u). 
 \end{equation}
The transformation law \eqref{27a}--\eqref{27b} is realized as follows
 \begin{equation}
 U(\Lambda)\varphi(x,u)U^{-1}(\Lambda)=\varphi(x',u').
 \end{equation}
Therefore the wave packets must satisfy the scalarity condition
 \begin{equation}
 f'(k',u')=f(k,u). 
 \end{equation}
It follows that the family $\{U(\Lambda)\}$ forms an unitary orbit of the 
intersystemic Lorentz group $L$ in the bundle of the Hilbert spaces 
$\mathcal{H}_{u}$; indeed we see that
 \begin{equation}
 \left(f',g'\right)_{u'}=\left(f,g\right)_{u}.
 \end{equation}
Now we introduce wave-packet solutions of the Klein--Gordon
equation \emph{via} the Fourier transformation
 \begin{equation}
 {\cal F}(x,u) = (2
\pi)^{-3/2} \int d\mu(k,u)\, f(k,u)e^{-ikx}.
 \end{equation}
 In terms of these solutions the scalar product (\ref{C2}) reads
 \begin{equation}\label{C6}
 ({\cal F},{\cal G})_{u} = - i \int d^{3}\vec{x}\, {\cal
F}^{\ast}(x,u) \overleftrightarrow{\partial}^0 {\cal
G}(x,u). 
 \end{equation}
 It is easy to see that for an orthonormal basis
$\{\Phi_{\alpha}(x,u)\}$ in $\mathcal{H}_{u}$ the completeness
relation holds
 \begin{equation}\label{C7}
 \sum_{\alpha} \Phi^{\ast}_{\alpha}(x,u) \Phi_{\alpha}(y,u) = i
\Delta^{+}_{T}(x - y,u),
 \end{equation}
 where $\Delta^{+}$ has the form (\ref{Q4}) and it is the reproducing
kernel in $\mathcal{H}_{u}$ i.e.\
 \[ (i \Delta^{+}(x,u), \Phi)_{u} = \Phi(x,u). \]

 Finally, the four-momentum operator has the form
 \begin{equation}\label{XXX1}
P_{\mu}=\int d\mu(k,u)\,k_{\mu} a^{\dagger}(k,u)
a(k,u).
 \end{equation}
Thus we have constructed a consistent quantum field theory for the hermitian, 
scalar tachyon field $\varphi(x, u)$. We conclude, that a proper framework 
to do this is the CT synchronization scheme.

\subsection{Spontaneous breaking of the synchronization group}\label{break}
As we have seen in the foregoing section, the intersystemic Lorentz
group $L$ is realized unitarily on the quantum level. In this section we
will analyse the role of the synchronization group $L_S$ in our scheme.

As was stressed in the Sec.\ \ref{syn}, if tachyons exist then one and
only one inertial frame is the preferred frame. In other words the
relativity principle is broken in this case: tachyons distinguish a
fixed synchronization scheme from the family of possible CT
synchronizations. Consequently, because all admissible synchronizations
are related by the group $L_S$, this group should be broken. To see this
let us consider transformations belonging to the subgroup $L_0$ (see
Sec.\ \ref{syn}). They are composed from the transformations of
intersystemic Lorentz group $L$ and the synchronization group $L_S$;
namely they have the following form (see eq.\ \eqref{L0} and the definition 
of $L_0$),
\begin{equation}\label{ZZ}
u' = u, \qquad x' = T(u) \Lambda_{0} T^{-1}(u) x \equiv \Lambda_{0}(u) x. 
\end{equation}
We search an operator $W(\Lambda_0)$ implementing (\ref{ZZ}) on the quantum 
level; namely
\begin{equation}\label{ZZZ}
\varphi'(x, u) = W(\Lambda^{-1}_{0}) \varphi(x, u) 
W^{\dagger}(\Lambda^{-1}_0) = \varphi(x', u).
\end{equation}
 This means that we should compare both sides of (\ref{ZZZ}) i.e.
\begin{multline}\label{84}
\int d\mu(k, u) \left[e^{i k x} a^{\prime\dagger}(k, u) + e^{-i k x} 
a^{\prime}(k, u)\right] \\
=\int d\mu(p, u) \left[e^{i p x'} a^{\dagger}(p, u) + e^{-i p x'} a(p, 
u)\right],
\end{multline}
where $x'$ is given by eq.\ (\ref{ZZ}), while, formally
\begin{equation}\label{85}
a' = W a W^{\dagger}, \qquad a^{\prime\dagger} = W a^{\dagger} W^{\dagger}.
\end{equation}
Taking into account the form of the measure $d\mu$ (eq.\ \eqref{measure}) and 
the fact that $\Lambda_{0}(u)$ does not leave invariant the sign of $k^0$, 
after some calculations, we deduce the following form of $W$:
\begin{gather}\label{86a}
a'(k,u)=\theta(k^{\prime 0})a(k',u)+
\theta(-k^{\prime 0})a^{\dagger}(-k',u),
\\ \label{86b}
a^{\prime\dagger}(k,u)=\theta(k^{\prime 0})a^{\dagger}(k',u)+
\theta(-k^{\prime 0})a(-k',u),
\end{gather}
where $k' = \Lambda^{-1}_0(u) k$.

We see that formally unitary operator $W(\Lambda_0)$ is realized by the
Bogolubov-like transformations; the Heaviside $\theta$-step functions
are the Bogolubov coefficients. The form (\ref{86a})--(\ref{86b}) of the
transformations of the group $L_0$ reflects the fact, that a possible
change of the sign of $k^0$ causes a different decomposition of the
field $\phi$ on the positive and negative frequencies. Furthermore it is
easy to check that the transformation (\ref{86a})--(\ref{86b}) preserves
the canonical commutation relations (\ref{aa1})--(\ref{aa+1}).

However, the formal operator $W(\Lambda_{0})$ realized in the ring of the
field operators, cannot be unitarily implemented in the space of states
in general; only if $\Lambda_0 = \Lambda_{u}$ is an element of the stability
group $SO(3)_u$ of $u$ in $L_S$, it can be realized unitarily. This is
related to the fact that $\Lambda_{u}$ does not change the sign of $k^0$ for 
any $k$. Indeed, notice firstly (see \eqref{86a}--\eqref{86b}) that 
$W(\Lambda_{0})$ does not anihilate the vacuum $\left|0\right>$. Moreover, 
the particle number operator
\begin{equation}\label{87}
N=\int d\mu(k,u) a^{\dagger}(k,u) a(k,u)
\end{equation}
applied to the ``new'' vacuum
\begin{equation} \label{88}
\left|0\right>'=W^{-1}\left|0\right>
\end{equation}
gives
\begin{equation}\label{89}
N\left|0\right>'=\delta^3(0)\int_{\Gamma}
d^3\nad{k}\,\theta(-(\Lambda_0(u) k)^0)\left|0\right>'.
\end{equation}
The right side of the above expression diverges like $\delta^6(0)$ for any 
$\Lambda_0(u) \in L_0/SO(3)_u$. Only for the stability subgroup $SO(3)_u$ 
vacuum remains invariant. Thus, a ``new'' vacuum $|0\rangle'$, related to an 
essentially new synchronization, contains an infinite number of ``old'' 
particles. As is well known, in such a case, two Fock spaces $H$ and $H'$, 
generated by creation operators from $|0\rangle$ and $|0\rangle'$ 
respectively, cannot be related by an unitary transformation\footnote{We can 
treat (\ref{86a})--(\ref{86b}), in some sense, as a quantum version of the 
familiar \emph{reinterpretation principle} \cite{Fei}. We find that the 
reinterpretation principle cannot be unitarily implemented---this is just the 
source of inconsistencies in approaches incorporating this principle.} 
($W(\Lambda_0)$ in our case). Therefore, we have deal with the so called 
spontaneous symmetry breaking of $L_S$ to the stability subgroup $SO(3)$ 
(recall that $L$ is realised unitarily). This means that physically 
privileged is only one realization of the canonical commutation relations 
(\ref{aa1})--(\ref{aa+1}) corresponding to the vacuum $|0\rangle$. Such a 
realization is related to a definite choice of the privileged inertial frame 
and consequently to a definite CT synchronization scheme. Thus we can 
conclude that, on the quantum level, \emph{tachyons distinguish a preferred 
frame via spontaneous breaking of the synchronization group}.

To complete discussion, let us apply the four-momentum operator
$P_{\mu}$ to the new vacuum $\left|0\right>'$. As the result we obtain
 \begin{equation}\label{90}
 P_{\mu}\left|0\right>'=-\delta^3(0){{\Lambda_0}_\mu}^{\nu}(u)
\int_{\Gamma}d^3\nad{k}\,\theta(-(\Lambda_0(u)k)^0)k_{\nu}
\left|0\right>'.
 \end{equation}
 This expression diverges again like $\delta^7(0)$ for $\Lambda_0\in
L_0/SO(3)_u$. Therefore a transition to a new vacuum ($\equiv$ change of
the privileged frame) demands an infinite momentum transfer, i.e.\ it is
physically inadmissible. This last phenomenon supports our claim that
existence of tachyons is associated with spontaneous breaking of the the
synchronization group.  On the other hand it can be shown
\cite{Rem2} that a free field theory for standard particles (bradyons or
luxons), formulated in CT synchronization, is unitarily equivalent to
the standard field theory in the EP synchronization; we do not repeat the 
corresponding proof because of its simplicity.

\boldmath
\section{Fermionic tachyons with helicity $\lambda=\pm\frac{1}{2}$}
\unboldmath
To construct tachyonic field theory describing field excitations with
the helicity $\pm\frac{1}{2}$, we assume that our field transforms under
Poincar\'e group like bispinor (for discussion of transformation rules
for local fields in the CT synchronization see \cite{Rem1}); namely
\begin{equation}
U(\Lambda) \psi(x, u) U(\Lambda^{-1}) = S(\Lambda^{-1}) \psi(x', u'),
\end{equation}
where $S(\Lambda)$ belongs to the representation
$D^{\frac{1}{2}0}\oplus D^{0\frac{1}{2}}$ of the Lorentz group. Because
we are working in the CT synchronization, it is convenient to introduce
an appropriate (CT-covariant) base in the algebra of Dirac matrices as
 \begin{equation}
 \gamma^{\mu}={T(u)^\mu}_{\nu}\gamma^{\nu}_{E},
 \end{equation}
 where $\gamma^{\mu}_{E}$ are standard $\gamma$-matrices, while $T(u)$
is given by the eq.\ (\ref{T:u}). Therefore
 \begin{equation}
 \{\gamma^{\mu},\gamma^{\nu}\}=2g^{\mu\nu}(u)I.
 \end{equation}
 However, notice that the Dirac conjugate bispinor 
$\bar\psi=\psi^\dagger\gamma^0_E$. Furthermore 
$\gamma^5=-\frac{i}{4!}\epsilon_{\mu\nu\sigma\lambda}
\gamma^\mu\gamma^\nu\gamma^\sigma\gamma^\lambda=\gamma^5_E$.

Now, we look for covariant field equations which are of degree one\footnote{In 
the Ref.\ \cite{Rem:neu} we found a class of the second order equations under 
condition of the $P$-invariance.}
with respect to the derivatives $\partial_{\mu}$ and imply the
Klein--Gordon equation
 \begin{equation}
 \left(g^{\mu\nu}(u)\partial_{\mu}\partial_{\nu}-\kappa^2\right)\psi=0,
 \end{equation}
 related to the space-like dispersion relation $k^{2}=-\kappa^{2}$. We also 
require the $T$-invariance of these equations.

As the result we obtain the following family of the Dirac-like equations
\begin{multline}
\left\{(u \gamma \sin\alpha - 1) \left((i u \partial) \cos\beta
-\kappa \sin\beta\right)-\gamma^5 \left[(-i \gamma \partial)
+\tfrac{i}{2} [\gamma \partial, u \gamma] \sin\alpha\right.\right.\\
\left.\left.+u \gamma \left((i u \partial)
(1 + \cos\alpha \sin\beta) + \kappa \cos\alpha
\cos\beta\right)\right]\right\}
\psi(x, u) = 0, \label{*}
\end{multline}
 derivable from an appriopriate hermitian Lagrangian density.
 Here $u\gamma=u_\mu\gamma^\mu$, $u\partial=u^\mu\partial_\mu$, 
$\gamma\partial=\gamma^\mu\partial_\mu$ and $\alpha$, $\beta$---real 
parameters, $\alpha\neq(2n+1)\frac{\pi}{2}$. To guarantee the
irreducibility of the elementary system described by (\ref{*}), the
equation (\ref{*}) must be  accompaniated by the {\em covariant\/}
helicity condition
\begin{equation}
\hat{\lambda}(u) \psi(u, k) = \lambda \psi(u, k) \label{**}
\end{equation}
where $\hat{\lambda}$ is given by (\ref{lam}) taken in the coordinate 
representation (see below) and $\lambda$ is fixed ($\lambda=\frac{1}{2}$ or 
$-\frac{1}{2}$ in our case). This condition is quite analogous to the 
condition for the left (right) bispinor in the Weyl's theory of the massless 
field. It implies that particles described by $\psi$ have helicity $-\lambda$, 
while antiparticles have helicity $\lambda$. For the obvious reason in the 
following we will concentrate on the case $\lambda=\frac{1}{2}$.

Notice that the pair of equations (\ref{*},\ref{**}) is not invariant under 
the $P$ or $C$ inversions separately for every choice of $\alpha$ and $\beta$.

Now, in the bispinor realization the helicity operator $\hat\lambda$ has the 
following explicit form
\begin{equation}
\hat\lambda(u)=\frac{\gamma^5 [i \gamma \partial, u \gamma]}
{4\sqrt{(i u \partial)^2 + \square}} \label{***}
\end{equation}
 where the integral operator 
$\left((i u \partial)^2 + \square\right)^{-\frac{1}{2}}$ in 
the coordinate representation is given by the well behaving distribution
\begin{equation}
\frac{1}{\sqrt{(-i u \partial)^2 + \square}} = \frac{1}{(2\pi)^4} \int 
\frac{d^4p\, \varepsilon(u p) e^{i p x}} {\sqrt{(u p)^2 - p^2}}. \label{****}
\end{equation}
 
Now, let us notice that the equation (\ref{*}), supplemented by the
helicity  condition (\ref{**}), are noninvariant under the composition
of the $P$ and $C$  inversions (see eqs.\ (\ref{Pa})--(\ref{Ca}) and the
Appendix), except of the  case $\sin\alpha=\cos\beta=0$. Because
(\ref{*})--(\ref{**}) are $T$-invariant,  therefore only for
$\sin\alpha=\cos\beta=0$ they are $CPT$-invariant. Taking
$\sin\beta=\cos\alpha=1$ we obtain from (\ref{*})
\begin{equation}\label{!}
\left\{\kappa + \gamma^5 [i \gamma \partial - 2 u \gamma (i u 
\partial)]\right\} \psi = 0,
\end{equation}
supplemented by (\ref{**}). On the other hand, for
$\cos\alpha=-\sin\beta=1$  we obtain
\begin{equation}\label{!!}
\left(\gamma^5(i\gamma\partial)-\kappa\right)\psi=0.
\end{equation}

The last equation is exactly the Chodos {\it et al}.\ \cite{CHK}
Dirac-like  equation for tachyonic fermion. However, contrary to the
standard EP approach,  it can be consistently quantized in our scheme
(if it is supplemented by the  helicity condition (\ref{**})). In the
following we will analyze the eqs.\  (\ref{!!}) and (\ref{**}) by means
of the Fourier decomposition
\begin{multline}
\psi(x, u) \\ = \frac{1}{(2\pi)^{\frac{3}{2}}}\int d^4k\,\delta(k^2+\kappa^2)
\theta(k^0)\left[w_{\frac{1}{2}}(k)e^{ikx}b^\dagger_{\frac{1}{2}}(k)
+v_{-\frac{1}{2}}(k)e^{-ikx}a_{-\frac{1}{2}}(k)\right] \label{*****}
\end{multline}
 of the field $\psi$. The creation and annihilation operators $a$ and $b$ 
satisfy the corresponding canonical anticommutation relations 
\eqref{aa}--\eqref{aa+}, i.e., the nonzero ones are
\begin{gather}\label{a1}
[a_{-\frac{1}{2}}(k),a^\dagger_{-\frac{1}{2}}(p)]_+
=2 \omega_{\mathbf{k}} \delta(\nad{k}-\nad{p})
\\ \label{a2}
[b_{\frac{1}{2}}(k),b^\dagger_{\frac{1}{2}}(p)]_+
=2 \omega_{\mathbf{k}} \delta(\nad{k}-\nad{p})
\end{gather}
 In (\ref{*****}) $b_{-\frac{1}{2}}$ and  $a_{\frac{1}{2}}$ do not
appear because we decided to fix $\lambda=\frac{1}{2}$ in (\ref{**}) 
(compare with (\ref{la})). As the consequence of (\ref{**}) the
corresponding  amplitudes $w_{-\frac{1}{2}}$ and $v_{\frac{1}{2}}$
vanish. The  nonvanishing amplitudes $w_{\frac{1}{2}}$ and
$v_{-\frac{1}{2}}$ satisfy
\begin{gather}\label{E1}
(\kappa+\gamma^5k\gamma)w_{\frac{1}{2}}(k,u)=0,
\\ \label{E2}
\left(1+\frac{\gamma^5 [k \gamma, u \gamma]}{2
\sqrt{q^2+\kappa^2}}\right)w_{\frac{1}{2}}(k,u)=0,
\\ \label{E3}
(\kappa-\gamma^5k\gamma)v_{-\frac{1}{2}}(k,u)=0,
\\ \label{E4}
\left(1+\frac{\gamma^5 [k \gamma, u \gamma]}{2
\sqrt{q^2+\kappa^2}}\right)v_{-\frac{1}{2}}(k,u)=0.
\end{gather} 
Here $k^0 = \omega_{\mathbf{k}}$, $q = u k$. The solution of 
(\ref{E1})--(\ref{E4}) 
reads
\begin{gather}\label{S1}
w_{\frac{1}{2}}(k,u)=\left(\frac{\kappa-\gamma^5k\gamma}{2\kappa}\right)
\frac{1}{2}\left(1-\frac{\gamma^5 [k \gamma, u \gamma]}{2
\sqrt{q^2+\kappa^2}}\right)w_{\frac{1}{2}}(\tilde{k}, \tilde{u}),
\\ \label{S2}
v_{-\frac{1}{2}}(k,u)=\left(\frac{\kappa+\gamma^5k\gamma}{2\kappa}\right)
\frac{1}{2}\left(1-\frac{\gamma^5 [k \gamma, u \gamma]}{2
\sqrt{q^2+\kappa^2}}\right)v_{-\frac{1}{2}}(\tilde{k}, \tilde{u}).
\end{gather}
Furthermore, the projections $w \bar{w}$ and $v \bar{v}$ read
\begin{gather}\label{w-w}
w_{\frac{1}{2}}(k, u) \bar{w}_{\frac{1}{2}}(k, u) = (\kappa-\gamma^5k\gamma)
\frac{1}{2}\left(1-\frac{\gamma^5 [k \gamma, u \gamma]}{2
\sqrt{q^2+\kappa^2}}\right),
\\ \label{v-v}
v_{-\frac{1}{2}}(k, u) \bar{v}_{-\frac{1}{2}}(k, u) = -(\kappa+\gamma^5k\gamma)
\frac{1}{2}\left(1-\frac{\gamma^5 [k \gamma, u \gamma]}{2
\sqrt{q^2+\kappa^2}}\right).
\end{gather}
The above amplitudes fulfil the covariant normalization conditions
\begin{gather}\label{N1}
\bar w_{\frac{1}{2}}(k,u)\gamma^5 u \gamma w_{\frac{1}{2}}(k,u)
=\bar v_{-\frac{1}{2}}(k,u)\gamma^5 u \gamma v_{-\frac{1}{2}}(k,u)
=2q,
\\ \label{N2}
\bar w_{\frac{1}{2}}(k^\pi,u)\gamma^5 u \gamma v_{-\frac{1}{2}}(k,u)
=0.
\end{gather} 
The amplitudes $w_{\frac{1}{2}}(\tilde{k}, \tilde{u})$ and 
$v_{-\frac{1}{2}}(\tilde{k}, \tilde{u})$, taken for the values $\tilde{k}$ 
and $\tilde{u}$ given in the eq.\ (\ref{ku}), have the following explicit 
form (for $\gamma^\mu_E$ matrix convention---see Appendix)
\begin{equation}\label{form}
w_{\frac{1}{2}}(\tilde{k}, \tilde{u}) = \begin{pmatrix} 0 \\ \sqrt{-q + 
\sqrt{q^2 + \kappa^2}} \\ 0 \\ \sqrt{q + \sqrt{q^2 + \kappa^2}} 
\end{pmatrix}, \quad
v_{-\frac{1}{2}}(\tilde{k}, \tilde{u}) = \begin{pmatrix} 0 \\ -\sqrt{-q + 
\sqrt{q^2 + \kappa^2}} \\ 0 \\ \sqrt{q + \sqrt{q^2 + \kappa^2}} 
\end{pmatrix}.
\end{equation}
It is easy to see that in the masseless limit $\kappa\to0$ the eqs.\ 
(\ref{E1})--(\ref{E4}) give the Weyl equations
\[ k\gamma w_{\frac{1}{2}}=k\gamma v_{-\frac{1}{2}}=0,\qquad
\gamma^5w_{\frac{1}{2}}=-w_{\frac{1}{2}},\quad
\gamma^5v_{-\frac{1}{2}}=-v_{-\frac{1}{2}}, \]
as well as the amplitudes (\ref{S1})--(\ref{S2}) have a smooth $\kappa\to0$ 
limit (it is enough to verify (\ref{form})).

Now, the normalization conditions (\ref{N1})--(\ref{N2}) generate the
proper work  of the canonical formalism. In particular, starting from
the Lagrangian density ${\cal
L}=\bar\psi\left(\gamma^5(i\gamma\partial)-\kappa\right)\psi$ we can 
derive the translation generators; with help of
(\ref{*****},\ref{a1},\ref{a2}) and (\ref{N1},\ref{N2}) we obtain
 \begin{equation}\label{Pm}
P_\mu=\int\frac{d^3\nad{k}}{2\omega_{\mathbf{k}}}k_\mu(
a^\dagger_{-\frac{1}{2}}a_{-\frac{1}{2}}+
b^\dagger_{\frac{1}{2}}b_{\frac{1}{2}})
\end{equation}
In agreement with (\ref{P:a}). Thus we have constructed fully consistent 
Poincar\'e covariant free field theory for a fermionic tachyon with helicity 
$-\frac{1}{2}$, quite analogous to the Weyl's theory for a left spinor which 
is obtained as the $\kappa \to 0$ limit.

\section{The stability of vacuum}\label{vacuum}
One of the serious defects of the standard approach to the tachyon field
quantization is apparent instability of the vacuum. The reason is that
relativistic kinematics admits in this case many-particle states with
vanishing total four-momentum. It is related directly to the fact that for
each (space-like) four-momentum, say $k^\mu_E$, the four-momentum
$-k^\mu_E$ with the opposite sign is kinematically admissible, because
there is no covariant spectral condition $k^0_E>0$ for space-like $k^\mu_E$.

Notwithstanding, such a situation does not take place in the presented 
scheme, because space-like four-momentum $k$ satisfies the invariant 
spectral condition, $k^0 > 0$ in each inertial frame\footnote{Recall that 
in the asymptotics $k^0 \to 0$ the wave packets decrease to zero (see 
remark below the eq.\ (\ref{XX3}).}. Thus the sum of $k$ and $k'$ satisfies 
the same spectral condition. In brief, we have exactly the same situation 
as in the case of the time-like (or light-like) four-momenta under the 
invariant spectral condition, $k^0_{\mathrm{E}}>0$. This means that in our 
scheme multiparticle states with vanishing total four-momentum do not 
appear, so the vacuum $|0\rangle$ cannot decay. For example, for two 
particle state $|q = k + p\rangle \equiv |k\rangle \otimes |p\rangle$, 
where $k$ and $p$ satisfy spectral condition, i.e., $k^0 > 0$, $p^0 > 0$, 
we have the inequality $q^0 > 0$ (i.e., $q \neq 0$), so there is no 
vacuum-like state with the four-momentum $q = 0$. Concluding, this theory 
is stable.

\section{Conclusions}
The main result of this work is demonstration that it is possible to 
construct Poincar\'e-covariant theory for tachyons on both classical and 
quantum level. The only price is necessity of existence of a preferred 
frame. Tachyons are classified according to the unitary representations of 
$SO(2)$ rather than $SO(2,1)$ group; so they are labelled by the 
eigenvectors of the helicity operator. In particular for the helicity 
$\lambda=\pm\frac{1}{2}$ we have constructed family  of $T$-invariant 
equations (\ref{*}). Under condition of $PCT$ invariance we  selected two 
equations (\ref{!}) and (\ref{!!}). The equation (\ref{!!})  coincide with 
the one proposed by Chodos \textit{et al}. \cite{CHK}. We show by  explicit 
construction that, in our scheme, theory described by this equation, 
supplemented by the helicity condition (\ref{**}) can be consistently 
quantized. This theory describe fermionic tachyon with helicity 
$-\frac{1}{2}$.  It has a smooth massless limit to the Weyl's left-handed 
spinor theory. These results show that there are no theoretical 
obstructions to interpret the  experimental data about square of mass of 
neutrinos \cite{PDG,Ass} as a signal that they can be fermionic tachyons. 
A more detailed discussion of this problem is given in the paper 
\cite{CibRem}.

We can conclude that, contrary to the current opinion, it is posible to agree 
the Lorentz covariance with universal causality and existence of a privileged 
frame. Moreover, a consistent quantization of the tachyonic field in this 
framework is possible. From this point of view the Einstein--Poincar\'e 
synchronization is useless in the tachyonic case---the proper choice is the 
CT synchronization. On the other hand, in a description of bradyons and 
luxons only, we are free in the choice of a synchronization procedure. For 
this reason we can use in this case CT-synchronization as well as the EP one. 

The CT-synchronization, a natural one for a description of tachyons, 
favourizes a reference frame (privileged frame). This preference is only 
formal (it is a convention) if tachyons do not exist. However, if they exist, 
then an inertial reference frame is really (physically) preferred, what in 
fact holds in the real world. As a consequence, the one-way light velocity 
can be measured in this case and, in general, it will be direction-dependent 
for a moving observer. Light velocity is isotropic only in the privileged 
frame. In the observed world we have a serious candidate to such a frame; 
namely frame related to the cosmic background radiation.

\appendix
\section{Derivation of the Lorentz group transformation rules}\label{der}
Let us derive the form of transformations between two coordinate frames 
$x^\mu$ and ${x'}^\mu$; for simplicity we denote $D(\Lambda, u(u_E)) \equiv 
\mathcal{D}(\Lambda, u_E)$
\begin{equation}\label{A1}
x'(u_E') = \mathcal{D}(\Lambda, u_E) x(u_E),
\end{equation}
where $\mathcal{D}(\Lambda, u_E)$ is a real (invertible) $4 \times 4$ matrix, 
$\Lambda$ belongs to the Lorentz group and $u_{E}^{\mu}$ is assumed to be a 
Lorentz four-vector, i.e.,
\begin{equation}\label{A2}
u_E' = \Lambda u_E,\quad {u_E}^2 = 1 >0.
\end{equation}
The transformations (\ref{A1})--(\ref{A2}) constitute a realization of the 
Lorentz group if the following composition law holds
\begin{equation}\label{D2}
\mathcal{D}({\Lambda}_2, {\Lambda}_1 u_E) \mathcal{D}({\Lambda}_1, u_E) = 
\mathcal{D}({\Lambda}_2 {\Lambda}_1, u_E).
\end{equation}
Now \emph{we demand that $(x^\mu) \equiv (x^0, \vec{x})$ transform
under subgroup of rotations as singlet + triplet (isotropy condition)}, i.e.\ 
for $R \in SO(3)$
\begin{equation}\label{A3}
\Omega \equiv \mathcal{D}(R, u_E) = \begin{pmatrix} 1 & 0 \\ 0 & R 
\end{pmatrix}.
\end{equation}
From eqs.\ (\ref{A1})--(\ref{D2}) we see that the identity and the inverse 
element have the form
\begin{gather}\label{A4a}
I = \mathcal{D}(I, u_E),\\
\label{A4b}
\mathcal{D}^{-1}(\Lambda, u_E) = \mathcal{D}(\Lambda^{-1}, \Lambda u_E).
\end{gather}
Using the familiar Wigner's trick we obtain that
\begin{equation}\label{A5}
\mathcal{D}(\Lambda, u_E) = T(\Lambda u_E) \Lambda T^{-1}(u_E),
\end{equation}
where the real matrix $T(u_E)$ is given by
\begin{equation}\label{A6}
T(u_E) = \mathcal{D}(L_{u_E}, \tilde{u}_E) L_{u_E}^{-1}.
\end{equation}
Here $\tilde{u}_E = (1, 0, 0, 0)$ and $L_{u_E}$ is the boost matrix: $u_E = 
L_{u_E} \tilde{u}_E$. We use the following parametrization of the matrix 
$L_{u_E}$
\begin{equation*}
L_{u_E} = \left(\begin{array}{c|c} u^0_E & \vec{u}_E^{\mathrm{T}} \\[1ex] 
\hline
\vec{u}_E & I + \frac{\vec{u}_E \otimes \vec{u}_E^{\mathrm{T}}}{(1 + u^0_E)}
\end{array}\right).
\end{equation*}

Note that the transformations (\ref{A1})--(\ref{A2}) leave the bilinear
form  $x^{\mathrm{T}}(u_E)\* g(u_E)\* x(u_E)$, where the symmetric tensor 
$g(u_E)$ reads
\begin{equation}\label{A7}
g(u_E) = (T(u_E) \eta T^{\mathrm{T}}(u_E))^{-1},
\end{equation}
invariant. Here $\eta$ is the Minkowski tensor and the superscript 
$^{\mathrm{T}}$ means transposition.

Now we determine the matrix $T(u_E)$. To do this we note that under rotations
\begin{equation*}
T(\Omega u_E) = \Omega T(u_E) \Omega^{-1}, 
\end{equation*}
so the most general form of $T(u_E)$ reads
\begin{equation}\label{A8}
 T(u_E)=\left(\begin{array}{c|c}
 a(u^0_E)&b(u^0_E)\vec u_E^{\rm T}\\[1ex]
 \hline
 d(u^0_E)\vec{u_E}& e(u^0_E)I+(\vec u_E\otimes\vec u_E^{\rm T}) f(u^0_E)
 \end{array}\right),
 \end{equation}
 where $a$, $b$, $d$, $e$ and $f$ are some functions of $u^0_E$.
Inserting eq.\ (\ref{A8}) into eq.\ (\ref{A7}) we can express the metric
tensor $g(u_E)$ by $a$, $b$, $d$, $e$ and $f$. In a three dimensional
flat subspace we can use an orthogonal frame (i.e.\
$(g^{-1})^{ik}=-\delta^{ik};$ $i,k=1,2,3$), so we obtain
 \begin{equation}\label{A9}
 e(u^0_E)=1,\quad d^2=(2-f\vec u_E^2)f.
 \end{equation}
 Furthermore, from the equation of null geodesics, $dx^{\rm T}g\,dx=0$,
we deduce that the light velocity $\vec c$ in the direction $\vec n$
($\vec n^2=1$) is of the form
\begin{equation}\label{A10}
\vec{c} = \vec{n} \left(\sqrt{\alpha + \beta^2 \vec{u}_E^2} - \beta \vec{u}_E 
\vec{n}\right)^{-1},
\end{equation}
where $\alpha=a^2-b^2\vec u_E^2$, $\beta=ad-b(1+f\vec u_E^2)$. From
eq.\ (\ref{A10}) we see that the synchronization convention depends on
the functions $\alpha$ and $\beta$ only. Now, because $a$, $b$ and $d$
can be expressed as functions of $\alpha$, $\beta$ and $f$ and we are
interested in {\em essentially different synchronizations\/} only, we can
choose
 \begin{equation}\label{A11a}
 f=0,
 \end{equation}
 so
 \begin{equation}\label{A11b}
 d=0,\quad\beta=-b,\quad\alpha=a^2-b^2\vec u_E^2.
 \end{equation}
 Finally, from (\ref{A10})--(\ref{A11b}) the average value of $|\vec c|$
over a closed path is equal to
\begin{equation*}
\langle|\vec{c}|\rangle_{\text{cl.\ path}} = \frac{1}{a}.
\end{equation*}
Because \emph{we demand that the round-trip light velocity}
($\langle|\vec c|\rangle_{\mbox{\scriptsize cl.\ path}} = c = 1$) {\em be
constant}, we obtain
 \begin{equation}\label{A12}
 a=1.
 \end{equation}
 Summarizing, $T(u_E)$ has the form
 \begin{equation}\label{A13}
 T(u_E)=\left(\begin{array}{c|c}
 1&b(u^0_E)\vec u_E^{\rm T}\\[1ex]
 \hline
 0&I\end{array}\right),
 \end{equation}
while the light velocity
 \begin{equation}\label{A14}
 \vec c = \vec n\left(1+b\vec u_E\vec n\right)^{-1},
 \end{equation}
so the Reichenbach coefficient reads
 \begin{equation}\label{A15}
 \varepsilon(\vec n,\vec u_E)=\frac{1}{2}\left(1+b\vec u_E\vec n\right).
 \end{equation}

In the special relativity the function $b(u^0_E)$ distinguishes between
different synchronizations. Choosing $b(u^0_E)=0$ we obtain $\vec{c} = 
\vec{n}$, $\varepsilon=\frac{1}{2}$ and the standard transformation
rules for coordinates: $x'_{E}=\Lambda x_{E}$, where, as before the
subscript~$_{E}$ denotes EP-synchronization. On the other hand, if we
demand that the instant-time hyperplane $x^0={\rm constant}$ be an
invariant notion, i.e.\ that ${x'}^0={\mathcal{D}(\Lambda, u_E)^0}_0 x^0$ so
${\mathcal{D}(\Lambda, u_E)^0}_k=0$, then from eqs.\ (\ref{A5}, \ref{A13}) we 
have
 \begin{equation}\label{A16}
 b(u^0_E)=-\frac{1}{u^0_E},
 \end{equation}
i.e. finally
\begin{equation}\label{A18a}
T(u_E) = \left(\begin{array}{c|c} 1 & 
-\frac{\vec{u}^{\mathrm{T}}_E}{u^0_E}\\[1ex] \hline 0 & I 
\end{array}\right).
\end{equation}
Thus we have determined the form of the transformation law (\ref{A1})
in this case in terms of the EP four-velocity $u_{E}$. Notice that in terms 
of $u = T(u_E) u_E$, the matrix $T = T(u)$, by means of eq.\ (\ref{A13}) has 
the form 
\begin{equation}
\left(\begin{array}{c|c} 1 & b(u) \vec{u}^{\mathrm{T}} \\ \hline 0 & I 
\end{array}\right)
\end{equation}
so for $b(u)$ determined by (\ref{A16}), $b = -u^0$, i.e.
\begin{equation}
T(u) = \left(\begin{array}{c|c} 1 & -u^0 \vec{u}^{\mathrm{T}} \\ \hline 0 & I 
\end{array}\right)
\end{equation}
in this case.

\section{Representation of the discrete symmetries}
The discrete transformations $P$, $T$ and $C$, defined by the eqs.\  
(\ref{Ps})--(\ref{Cs}) are realised in the bispinor space standardly, i.e.  
$P$ by $\gamma^0_E$, while $T$ and $C$ by $\mathcal{T}$ and $\mathcal{C}$ 
satisfying the conditions
\begin{gather}\label{Tc}
\mathcal{T}^\dagger \mathcal{T} = I, \quad \mathcal{T}^* \mathcal{T} = -I, 
\quad \mathcal{T}^{-1} {\gamma^\mu}^{\mathrm{T}} \mathcal{T} = \gamma^\mu,
\\ \label{Cc}
\mathcal{C}^\dagger \mathcal{C} = I, \quad \mathcal{C}^* \mathcal{C} = -I, 
\quad \mathcal{C}^{\mathrm{T}} = -\mathcal{C}, \quad \mathcal{C}^{-1} 
\gamma^\mu \mathcal{C} = -{\gamma^\mu}^{\mathrm{T}}.
\end{gather}
Notice that the last condition in (\ref{Tc}) and (\ref{Cc}) can be 
fortmulated in terms of  the standard $\gamma^\mu_E$ exactly in the same 
form.

In explicit calculations of the amplitudes (\ref{form}) we have used the 
following representations of the $\gamma_E$ matrices: $\vec{\gamma}_E = 
\begin{pmatrix} 0 & -\vec{\sigma} \\ \vec{\sigma} & 0\end{pmatrix}$, 
$\gamma^0_E = \begin{pmatrix} 0 & I \\ I & 0 \end{pmatrix}$. In this 
representation the parity, charge conjugation and time inversion are given, 
up to a phase factor by
\begin{equation*}
\mathcal{P} = \gamma^0_E, \quad \mathcal{C} = i \gamma^0_E \gamma^2_E, \quad 
\mathcal{T} = -i \gamma^0_E \gamma^2_E \gamma^5_E.
\end{equation*}
Finally, we use $\epsilon^{\mu\nu\sigma\lambda}$ with $\epsilon^{0123} = 1$; 
consequently $\gamma^5_E = i \gamma^0_E \gamma^1_E \gamma^2_E \gamma^3_E$.


\end{document}